\documentclass[fleqn,10pt]{wlscirep}
\usepackage[utf8]{inputenc}
\usepackage[T1]{fontenc}
\usepackage{url}
\usepackage{graphicx}
\usepackage{booktabs}       
\usepackage{amsfonts}       
\usepackage{amsmath,amsfonts,amssymb}
\usepackage{bm}
\usepackage{upgreek}
\usepackage{mathtools}
\usepackage{siunitx}
\usepackage{nicefrac}       
\usepackage{microtype}      
\usepackage{xcolor,soul}         
\usepackage{nameref}
\usepackage{float}
\usepackage{lineno}

\newcommand{\shortmethodname}{RINGER}
\newcommand{\longmethodname}{RINGER Generates Ensembles of Rings}

\makeatletter
\newcommand{\@giventhatstar}[2]{\ensuremath{\left(#1\;\middle|\;#2\right)}}
\newcommand{\@giventhatnostar}[3][]{#1(#2\;#1|\;#3#1)}
\newcommand{\giventhat}{\@ifstar\@giventhatstar\@giventhatnostar}
\makeatother

\DeclarePairedDelimiterX{\infdivx}[2]{(}{)}{%
  #1\;\delimsize\|\;#2%
}

\DeclareSIUnit\angstrom{\text {Å}}

\DeclareMathOperator*{\argmin}{arg\,min} 

\title{Accurate and Efficient Structural Ensemble Generation of Macrocyclic Peptides using Internal Coordinate Diffusion}

\author[1*]{Colin A. Grambow}
\author[1]{Hayley Weir}
\author[2]{Nathaniel L. Diamant}
\author[2]{Gabriele Scalia}
\author[2]{Tommaso Biancalani}
\author[1*]{Kangway V. Chuang}

\affil[1]{Prescient Design, Genentech, 1 DNA Way, South San Francisco, CA, 94080}
\affil[2]{Department of Biological Research and Artificial Intelligence Development, Genentech, 1 DNA Way, South San Francisco, CA, 94080}

\affil[*]{corresponding authors: Colin A. Grambow (grambow.colin@gene.com) and Kangway V. Chuang (chuang.kangway@gene.com)}

\begin{abstract}
Macrocyclic peptides are an emerging therapeutic modality, yet computational approaches for accurately sampling their diverse 3D ensembles remain challenging due to their conformational diversity and geometric constraints. Here, we introduce \shortmethodname, a diffusion-based transformer model using a redundant internal coordinate representation that generates three-dimensional conformational ensembles of macrocyclic peptides from their 2D representations. \shortmethodname{} provides fast backbone and side-chain sampling while respecting key structural invariances of cyclic peptides. Through extensive benchmarking and analysis against gold-standard conformer ensembles of cyclic peptides generated with metadynamics, we demonstrate how \shortmethodname{} generates both high-quality and diverse geometries at a fraction of the computational cost. Our work lays the foundation for improved sampling of cyclic geometries and the development of geometric learning methods for peptides. 
\end{abstract}

\begin{document}

\flushbottom
\maketitle
\thispagestyle{empty}

\section*{Introduction}

Macrocyclic peptides are an important therapeutic modality in modern drug discovery that occupy a unique chemical and pharmacological space between small and large molecules~\cite{Driggers2008-ax, Muttenthaler2021-jw, Vinogradov2019-bl}. These cyclic peptides exhibit improved structural rigidity and metabolic stability compared to their linear counterparts~\cite{Craik2013-od}, yet retain key conformational flexibility and diversity to bind shallow protein interfaces~\cite{Villar2014-uf}. However, computational approaches for predicting their structural ensembles remain limited compared to small molecules and proteins in terms of computational speed, accuracy (sample quality), and conformational diversity~\cite{Poongavanam2018-oo}. Critically, scalable and accurate tools for predicting conformational ensembles are necessary to enable rational design of macrocyclic drugs; access to these tools can significantly impact optimization of key properties including binding affinity~\cite{Alogheli2017-zf, Garcia_Jimenez2023-ej}, permeability~\cite{Leung2016-mc, Rezai2006-us, Bhardwaj2022-jy, Mulligan2020}, and oral bioavailability~\cite{Nielsen2017-yz}.

Recent work has expanded the scope and diversity of macrocycles that can be computationally designed and experimentally validated~\cite{Hosseinzadeh2017,Salveson2024}. These advances enable exploring a vast space of drug-like macrocycles to enhance structure-based drug design. However, a key challenge remains in predicting the conformational ensembles of macrocycles, as they often adopt multiple relevant conformations to balance drug-like properties such as permeability and binding affinity~\cite{Bhardwaj2022-jy}. Emerging methods in the field of protein structure prediction are pushing the frontiers in sampling alternative conformational states~\cite{Wayment-Steele2024}. Continued development and adaptation of such approaches for macrocycles could provide valuable insights into the role of conformational flexibility in modulating their biological activity and facilitating the design of effective macrocyclic therapeutics.

Several key challenges hinder fast and effective macrocycle structure generation: 1) Macrocyclic peptides exhibit diverse molecular structures and chemical modifications, including varying ring size, stereochemistry, \textit{N}-methylation, and more~\cite{Kamenik2018-ui}.  Their structural diversity, along with the increased number of rotatable bonds, results in a vast conformational space that is considerably more expensive to sample computationally. 2) Macrocycles are subject to complex non-linear constraints due to ring closure. The atomic positions, angles, and dihedrals of the macrocycle backbone are highly interdependent, and additional complex intramolecular interactions make this process inherently difficult to model~\cite{Watts2014-no}. 3) Experimental X-ray and NMR structures for macrocycles are lacking ($\sim$\num{e3}) in comparison to small molecules ($\sim$\num{e6} in the Cambridge Structural Database~\cite{Groom2016-us}) and proteins ($\sim$\num{e5} in the Protein Data Bank~\cite{Berman2000-wt}). The scarcity of available experimental data has made it difficult to integrate observational data to improve structural predictions or train machine learning-based approaches. Together, the vast conformational space combined with limited data make modeling and sampling of macrocycles not only conceptually challenging, but technically challenging due to computational cost. Existing approaches cannot accurately generate diverse and accurate conformational ensembles at a scale necessary to accelerate and enable macrocycle design.

\begin{figure}[t]
    \centering
    \includegraphics[width=1.0\linewidth]{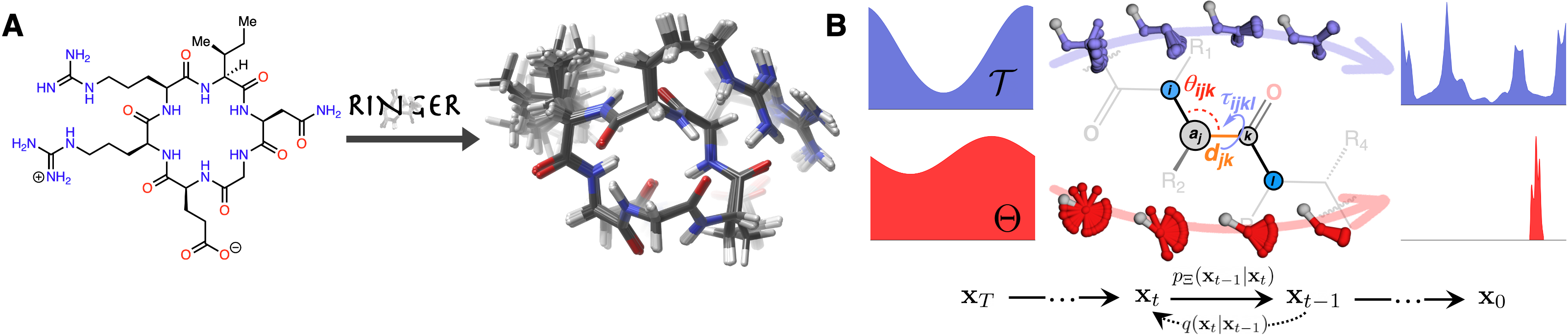}
    \vskip -0.05in
    \caption{Overview of \shortmethodname{} for macrocycle conformer generation. {\bf A.} Given a 2D representation of a macrocyclic peptide, \shortmethodname{} generates an accurate and diverse 3D conformational ensemble. {\bf B.} An illustration of the diffusion process learning to recover the time $t=0$ bond angle (red) and torsional (blue) distributions from time, $t=T$.}
    \label{fig:overview}
\end{figure}

To address these limitations, we introduce \shortmethodname{} (\longmethodname), a deep learning model designed specifically for sequence-conditioned macrocycle structure generation (Figure~\ref{fig:overview}) that efficiently samples realistic angles and torsions (i.e., internal coordinates) for macrocyclic peptides. \shortmethodname{} merges a transformer architecture that naturally captures the physical equivariances and invariances of macrocyclic peptides with a discrete-time diffusion model to learn highly-coupled distributions over internal coordinates. We demonstrate how \shortmethodname{} simultaneously achieves excellent performance in sample quality over angular and torsional profiles while maintaining excellent RMSDs relative to gold-standard conformer ensembles generated with the Conformer-Rotamer Ensemble Sampling Tool (CREST)~\cite{Pracht2020-ha, CREST2024}. 

We summarize our contributions as follows:

\begin{itemize}
    \item{We propose a new framework, \shortmethodname{}, for all-atom conformer generation of macrocycles based on efficiently encoding their geometry using redundant internal coordinates. Our model naturally handles the cyclic nature of macrocycles and chiral side chains with both \textsc{L}- and \textsc{D}-amino acids, and we propose a simple solution to satisfy ring-closure constraints.}
    \item{We benchmark \shortmethodname{} extensively against state-of-the-art physics- and machine learning-based algorithms to demonstrate how our approach better captures complex and highly coupled geometric distributions. RINGER outperforms state-of-the-art algorithms that predict 3D conformer ensembles from their 2D structures for macrocyclic peptides.}
    \item{Our work highlights the advantages of a machine learning-based approach for structure generation in terms of quality, diversity, and  speed. We show how RINGER predicts diverse and complete conformational ensembles with excellent sample quality that nearly match the gold-standard ensembles generated with expensive metadynamic simulations.}
\end{itemize}

\section*{Background and Related Work}
Our work builds on small-molecule conformer generation and protein structure modeling to create a framework for macrocycle conformers. Below, we briefly summarize related work.
\vspace{-0.2cm}

\paragraph{Physics and Heuristics-based Conformer Generation for Macrocycles}
Physics-based and heuristics-based algorithms remain the state of the art for macrocycles and have required special considerations compared to drug-like small molecules due to ring-closing constraints. The open-source cheminformatics library RDKit leverages distance geometry algorithms for small-molecule conformer generation (ETKDG)~\cite{Riniker2015-kk}, with improved heuristic bounds for macrocycles (ETKDGv3)~\cite{Wang2020-ai, Wang2022-hb}. Similarly, commercial software such as OpenEye OMEGA~\cite{Hawkins2010-xq, Hawkins2012-aj} in \texttt{macrocycle} mode uses a distance geometry algorithm based on 4D coordinate initialization to provide diverse conformers~\cite{Spellmeyer1997-wx}, as their torsion-driving approach is incompatible with ring closure.

Similarly, low-mode~\cite{Kolossvary1996-mo,Kolossvry1999-rf} or Monte Carlo~\cite{Chang1989-fp} search methods combined with molecular dynamics have been found to be effective at sampling macrocycle conformations, particularly when combined with force field optimizations as demonstrated in Schr\"odinger's MacroModel~\cite{Watts2014-no} and Prime MCS~\cite{Sindhikara2017-te}. These approaches have been tuned with expert knowledge and torsional libraries to maximize agreement with observed experimental structures. The open-source CREST package~\cite{Pracht2020-ha,CREST2024} leverages multi-start iterative metadynamics with a genetic structure-crossing algorithm (iMTD-GC) to explore new geometries, and can be considered a gold-standard for generating diverse ensembles of drug-like molecules. In this work, we use the recently-published CREMP~\cite{grambow023cremp} dataset, containing high-quality, CREST-generated ensembles, representing over 31 million macrocycle geometries (see the \nameref{dataset} section and Supplementary Section~\ref{app:dataset} for more details). 

One key limitation of these approaches is high computational cost and difficulty in scaling; in general, conformer generation is \numrange{e3}{e5} times more computationally expensive compared to a drug-like small molecule due to the increased number of rotatable bonds and their ring-closing constraints (e.g., generating a conformational ensemble of a macrocyclic hexapeptide with CREST requires an average of 14~hours~\cite{grambow023cremp}). These approaches become increasingly challenging when kinetic or molecular dynamics approaches are used with explicit solvation~\cite{Damjanovic2021-qx, Linker2023-sp}.

\paragraph{Generative Approaches for Small Molecule Conformer Ensembles} 
Recent work with deep generative models has focused on improved sampling of the conformational landscape of small molecules. For example, Mansimov et al.~\cite{Mansimov2019-sc} propose a conditional graph variational autoencoder (CGVAE) approach for molecular geometry generation. Simm \& Hernandez-Lobato~\cite{pmlr-v119-simm20a} report conditional generation of molecular geometries based on distance geometry. Xu et al.~\cite{xu2021learning} leverage normalizing flows and energy-based modeling to help capture the multimodal nature and complex dependencies of small molecule space. More recently, Xu et al.~\cite{xu2022geodiff} report GeoDiff, an equivariant diffusion-based model that operates on Cartesian point clouds. Although GeoDiff provides strong results, sampling is costly and requires 5,000 time steps. Recently, Zhu et al.\cite{zhu2022dmcg} report DMCG, which leverages a variational autoencoder to directly predict coordinates while maintaining invariance to rototranslation and permutation of symmetric atoms.

Recent reports have also drawn inspiration from physics-based conformer generation to leverage the rigid-rotor hypothesis, which treats bond distances and angles as fixed, and torsional angles of rotatable bonds are independently sampled, assuming little or no interdependence between torsions~\cite{Scharfer2013-or}. These include GeoMol~\cite{ganea2021geomol}, an SE(3)-invariant machine learning model for small molecule conformer generation that leverages graph neural networks, and EquiBind~\cite{pmlr-v162-stark22b} which performs conditional generation on protein structure. Recently, Jing et al.~\cite{Jing2022-torsionaldiffusion} report Torsional Diffusion, a diffusion model that operates on the torsional space via an extrinsic-to-intrinsic score model to provide strong benchmarks on the GEOM dataset~\cite{Axelrod2022-mw}. Importantly, these methods do not address the challenge of highly-coupled torsions within cyclic systems and either propose complex ring-averaging processes~\cite{ganea2021geomol} or disregard sampling of cyclic structures all together~\cite{Jing2022-torsionaldiffusion}.

\paragraph{Protein Structure Prediction and Diffusion} Significant progress has been made recently in protein structure prediction with the advent of methods such as AlphaFold2~\cite{Jumper2021-fg} and RoseTTAFold~\cite{Baek2021-kv}. However, structure prediction methods have predominantly focused on deterministic maps to static output structures rather than on sampling diverse structure ensembles. Recently, several papers have developed diffusion-based approaches for protein generation based on Euclidean diffusion over Cartesian coordinates~\cite{anand2022protein, yim2023se3} or backbones as in FoldingDiff~\cite{wu2022protein}, with an emphasis on structural design. FoldingDiff parameterizes structures over internal backbone angles and torsions and relies on the natural extension reference frame (NeRF)~\cite{Parsons2005-yj} to perform linear reconstructions. However, as we demonstrate below, naive linear reconstructions fail to address the ring constraints for macrocycles. Moreover, FoldingDiff focuses on \emph{unconditional} generation of protein backbones and is hence not suitable for all-atom conformer ensemble generation. Below, we focus on the challenging problem of conditional generation of constrained geometries.
 
\paragraph{Machine Learning Approaches for Macrocycle Conformer Ensemble Generation} Despite the many approaches focused on small molecules and protein structure generation, there are few efforts in macrocycle structure prediction. Most notably, Miao et al.~\cite{Miao2021-fj} recently disclosed StrEAMM for learning on molecular dynamics of cyclic peptides using explicit solvation. StrEAMM is a linear model that predicts local backbone geometries and their respective 1,2- and 1,3-residue interactions to provide excellent ensemble estimates of homodetic hexapeptides, but does not generate explicit all-atom conformers. Additionally, the model is not naturally inductive and is not natively extensible to other macrocycle ring sizes and residues. Fishman et al.~\cite{fishman2023diffusion} recently developed a more general framework for diffusion models on manifolds defined via a set of inequality constraints. However, they only investigate the conformational ensemble of a single cyclic peptide as a proof-of-concept using a reduced $\alpha$-carbon representation.

\section*{\shortmethodname: Problem Statement and Methods}

\subsection*{Problem Definition: Conditional Macrocycle Conformer Generation}

The core objective of our work is to model the distribution of conformers for a macrocyclic peptide with a given amino acid sequence. Given a peptide macrocycle graph $\mathcal{G} = \mathcal{(V, E)}$, where $\mathcal{V}$ is the set of nodes (atoms) and $\mathcal{E}$ is the set of edges (bonds), and $n = |\mathcal{V}|$ our goal is to learn a distribution over the possible conformers. Let $\mathcal{C} = \{c_1, c_2, \dots, c_K\}$ be the set of conformers, where each conformer $c_k \in \mathcal{C}$ represents a unique spatial arrangement of the atoms $\mathcal{V}$. Our task is to learn the distribution $p\giventhat{\mathcal{C}}{\mathcal{G}}$, which represents the probability over the conformer ensemble $\mathcal{C}$ given a molecular graph $\mathcal{G}$. Learning and sampling from this complex distribution is inherently challenging for most molecules, and is further complicated in macrocycles due to the highly-coupled nature of ring atoms. A perturbation to one part of the ring generally perturbs the others. Consequently, any model must account for the interdependence between atoms due to the cyclic constraints.

Given this problem, a good generative model ideally satisfies a few key properties: 1) Naturally encodes the physical and structural aspects of macrocyclic peptides. For example, cyclic peptides with only standard peptide bonds (i.e., homodetic peptides) do not have a natural starting residue and hence exhibit cyclic shift invariance, e.g., cyclo-(R.I.N.G.E.R) is identical to cyclo-(I.N.G.E.R.R), where each amino acid is denoted by its one-letter code with ``cyclo'' indicating cyclization of the sequence. 2) Captures multimodal distributions and complex, higher-order interactions such as the strong coupling between atomic positions in the ring. 3) Samples high-quality and diverse conformations from $p\giventhat{\mathcal{C}}{\mathcal{G}}$ that faithfully capture realistic geometries while respecting the underlying conformer distribution.

\subsection*{Representing Macrocycle Geometry: Redundant Internal Coordinates}

Conformer geometries are defined by their set of Cartesian coordinates for each atomic position and can hence be modeled using SE(3)-equivariant models to learn complex distributions. However, Euclidean diffusion requires modeling the many degrees of freedom; and, in practice, can require many time steps to generate accurate geometries~\cite{xu2022geodiff}. Moreover, realistic conformations are highly sensitive to the precise interatomic distances, angles, and torsions---although this information is implicit in the Cartesian positions, explicitly integrating these quantities into a model can provide a strong inductive bias and accelerate learning~\cite{Gasteiger_undated-za}.

Borrowing from molecular geometry optimization~\cite{peng1996}, protein representation~\cite{Ramachandran1968-mn,dunbrack1994,Parsons2005-yj}, and inverse kinematics~\cite{Han2006InverseKF}, we adopt redundant internal coordinates that represent conformer geometries through a set of bond distances, angles, and torsions (dihedral angles), i.e., $\mathcal{C} \equiv \{\mathcal{D}, \Theta, \mathcal{T}\}$. In particular, this simplifies the learning task, as bond distances can be approximated as fixed distances with little loss in accuracy~\cite{Hawkins2010-xq,wu2022protein,Jing2022-torsionaldiffusion}, and internal angles typically fit a narrow distribution. Importantly, these coordinates define an internal reference frame that readily encodes complex geometries including chirality. Moreover, this approach obviates the need for complex equivariant networks~\cite{xu2022geodiff, Jing2022-torsionaldiffusion}. Hence, our generative process can be reformulated as learning the distribution $p\giventhat{\{\Theta, \mathcal{T}\}}{\mathcal{G};\mathcal{D}}$ using fixed bond distances for reconstruction back to Cartesians (Figure~\ref{fig:overview}).

\subsection*{Deep Probabilistic Diffusion Models for Sampling Internal Coordinates}

\paragraph{Denoising Probabilistic Models} 
Recent works on deep denoising probabilistic models have demonstrated excellent generative performance for complex multimodal data~\cite{pmlr-v37-sohl-dickstein15, Ho-denoising-2020, song2021scorebased}, and have been successfully applied to both small molecules and proteins~\cite{xu2022geodiff, wu2022protein}. We adapt a discrete-time diffusion process~\cite{wu2022protein} that formulates the forward transition probability using a wrapped normal distribution, $q\giventhat*{\mathbf{x}_t}{\mathbf{x}_{t-1}} = \mathcal{N}_{\text {wrapped}}\left(\mathbf{x}_t ; \sqrt{1-\beta_t} \mathbf{x}_{t-1}, \beta_t \mathbf{I}\right)$, instead of a standard normal distribution~\cite{Jing2022-torsionaldiffusion}, where $\mathbf{x}_t$ represents the noised internal coordinates (bond angles and torsions) at time step $t$. We train a diffusion model, $p_{\Xi}\giventhat{\mathbf{x}_{t-1}}{\mathbf{x}_t}$, by training a neural architecture, described below, to predict the noise present at a given time step (for full details, see Supplementary Section~\ref{app:training}). During inference, we sample $\mathbf{x}_T$ from a wrapped normal distribution and iteratively generate $\mathbf{x}_0$ using $p_{\Xi}\giventhat{\mathbf{x}_{t-1}}{\mathbf{x}_t}$. The sampling process is further detailed in Supplementary Section~\ref{app:sampling}.

\paragraph*{Encoder Architecture}
Macrocycles exhibit extensive coupling of their residues due to torsional strain and intramolecular interactions such as hydrogen bonds. Here, we use self-attention~\cite{vaswani_transformer,devlin-etal-2019-bert} to learn the complex interactions between atoms. Unlike standard sequence models for linear data, macrocycles exhibit cyclic symmetry with no canonical start position. Thus, we design a modified bidirectional, relative positional encoding~\cite{shaw-etal-2018-self} $\mathbf{p}_{ij}^K$ to reflect this physical invariance (see Supplementary Section~\ref{app:glossary} for notation and \ref{app:cyc_ablation} for an ablation study):

\begin{equation}
\mathbf{z}_i=\sum_{j=1}^n \alpha_{i j}\left(\mathbf{v}_j \mathbf{W}^V\right), \quad \text{where} \quad \alpha_{i j}=\frac{\exp e_{i j}}{\sum_{k=1}^n \exp e_{i k}}
\label{eq:cyclic1}
\end{equation}

\begin{equation}
e_{i j}=\frac{\mathbf{v}_i \mathbf{W}^Q\left(\mathbf{v}_j \mathbf{W}^K+\mathbf{p}_{i j}^K\right)^T}{\sqrt{d_z}}  \quad \text{with} \quad \mathbf{p}_{ij}^K = 
\underbrace{\mathbf{W}^{D}_{\left(i-j \right) \bmod n}}_{\text{forward}}
+  
\underbrace{\mathbf{W}^{D}_{\left(i-j \right) \bmod (-n)} }_{\text{backward}}
\label{eq:cyclic2}
\end{equation}

\noindent These cyclic relative position representations specify forward (N-to-C) and reverse (C-to-N) relationships between each atom in the macrocycle, and effectively encode the molecular graph in a self-attention module similar to Ying et al.~\cite{graphormer2021}. The relative position of any neighboring atom is uniquely defined by its forward and reverse graph distances in the embedding lookup $\mathbf{W}^D$. For conditional generation, we perform a linear projection of the atom features $\mathbf{a}_i$, corresponding to \emph{each macrocycle backbone atom and its side chain}, and a separate linear projection of the angles and torsions $\mathbf{x}_i = \bm{\uptheta}_i \oplus \bm{\uptau}_i$ and concatenate them as a single input to the transformer, $\mathbf{v}_i = \mathbf{a}_i' \oplus \mathbf{x}_i'$. Notably, our diffusion model only adds noise to the angular component, $\mathbf{x}_i$, corresponding to backbone and side-chain angles and torsions that define a complete atomic configuration. For unconditional backbone generation, atoms are only labeled with their backbone identity (nitrogen, $\alpha$-carbon, carbonyl-carbon) using an embedding that is added to the input, and side chains are not modeled. Model details are shown in Supplementary Section~\ref{app:model}.

\paragraph{Ring Closing: Back Conversion to Cartesian Ring Coordinates} Macrocycles with fixed bond distances contain three redundant torsional angles and two redundant bond angles. Whereas linear peptides and proteins can be readily converted into an arbitrary Cartesian reference frame through methods such as NeRF~\cite{Parsons2005-yj}, these redundancies prevent direct transformation to unique Cartesians for cyclic structures. Adopting a sequential reconstruction method such as NeRF accumulates small errors that result in inadequate ring closure for macrocycles.\footnote{Although direct equality and inequality constraints over the diffusion process is a promising direction that could address this problem, we leave this for future work.} Other studies have developed complex heuristics with coordinate averaging for ring smoothing~\cite{ganea2021geomol}, yet these approaches can distort the predicted geometries. In practice, we demonstrate that an efficient post-processing step works well with minimal distortion: we treat this as a constrained optimization problem using the Sequential Least Squares Quadratic Programming (SLSQP) algorithm~\cite{slsqp-kraft1988} to ensure valid Cartesian coordinates while satisfying distance constraints:

\begin{equation}
    \bm{\hat{\upxi}} = \argmin_{\bm{\upxi}}\ \lVert \bm{\uptheta}(\bm{\upxi}) - \bm{\hat{\uptheta}} \rVert^2 + \lVert w\left(\bm{\uptau}(\bm{\upxi}) - \bm{\hat{\uptau}} \right) \rVert^2 \quad \text{subject to:} \quad \
    \mathbf{d}(\bm{\upxi}) = \mathbf{d}_{\text{true}}
\label{eq:opt}
\end{equation}

\noindent Here, we find the set of \emph{ring} Cartesian coordinates, $\bm{\hat{\upxi}}$, that minimize the squared error against the ring internal coordinates $\bm{\hat{\uptheta}}$ and $\bm{\hat{\uptau}}$ sampled by the diffusion process while satisfying bond distance equality constraints using known ring bond distances, $\mathbf{d}_{\text{true}}$, from the training data. The torsion error, $\bm{\uptau}(\bm{\upxi}) - \bm{\hat{\uptau}}$, is wrapped by $w(\cdot)$ so that it remains in the $[-\pi,\pi)$ range. Empirically, we demonstrate that this scheme recovers realistic macrocycles with high fidelity by evenly distributing the error across the entire macrocycle backbone (see Supplementary Section~\ref{app:opt} for additional details). Additionally, we reject a sample from \shortmethodname{} if the ring torsion fingerprint deviation~\cite{Wang2020-ai} before versus after optimization with Equation~\eqref{eq:opt} is larger than 0.01, which further improves the quality of generated ensembles.

\paragraph{Overall Generation Procedure} Our model represents macrocycles as cyclic sequences of backbone atoms with fixed bond lengths, where each atom is featurized with two ring internal coordinates and several side chain internal coordinates. We train a discrete-time diffusion model to learn a denoising process over the internal coordinates using a transformer architecture with an invariant cyclic positional encoding. Chemical features of the backbone atoms and molecular fingerprints of the attached side-chain atoms provide 2D macrocycle structure information for the sequence-conditioned models. At inference time, we sample from a wrapped Gaussian distribution to produce a set of angles and torsions. In the final post-processing step, macrocycle geometries with Cartesian coordinates are reconstructed through our constrained optimization using Equation~\eqref{eq:opt} for ring atoms and using NeRF~\cite{Parsons2005-yj} for rotatable side-chain atoms. Hydrogens and non-rotatable side-chain groups like phenyl rings are not modeled directly and are generated using RDKit. A detailed flowchart illustrating the main steps of our method is shown in Supplementary Section~\ref{app:overview}.

\section*{Results}

\subsection*{Unconditional Generation of Macrocycle Backbones}

\begin{figure}[tb]
    \centering
    \begin{minipage}{0.71\linewidth}
        \includegraphics[width=\linewidth]{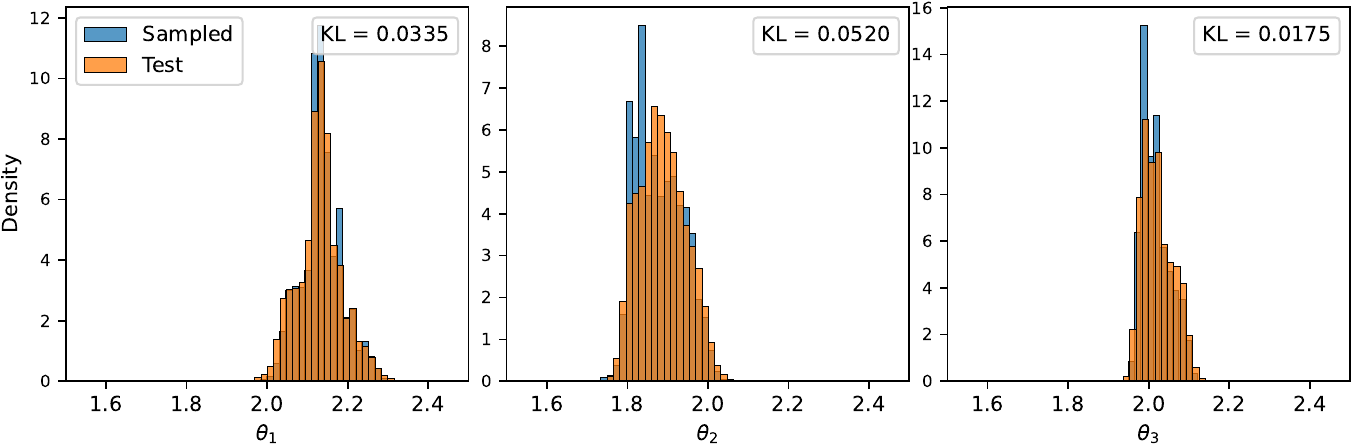}
        \includegraphics[width=\linewidth]{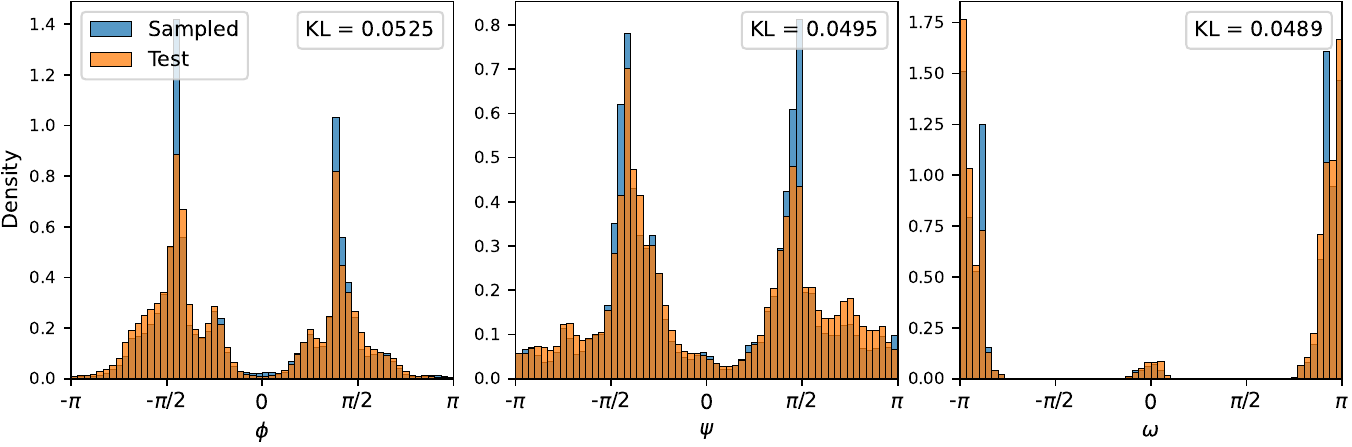}
    \end{minipage}\hfill%
    \begin{minipage}{0.275\linewidth}
        \includegraphics[width=\linewidth]{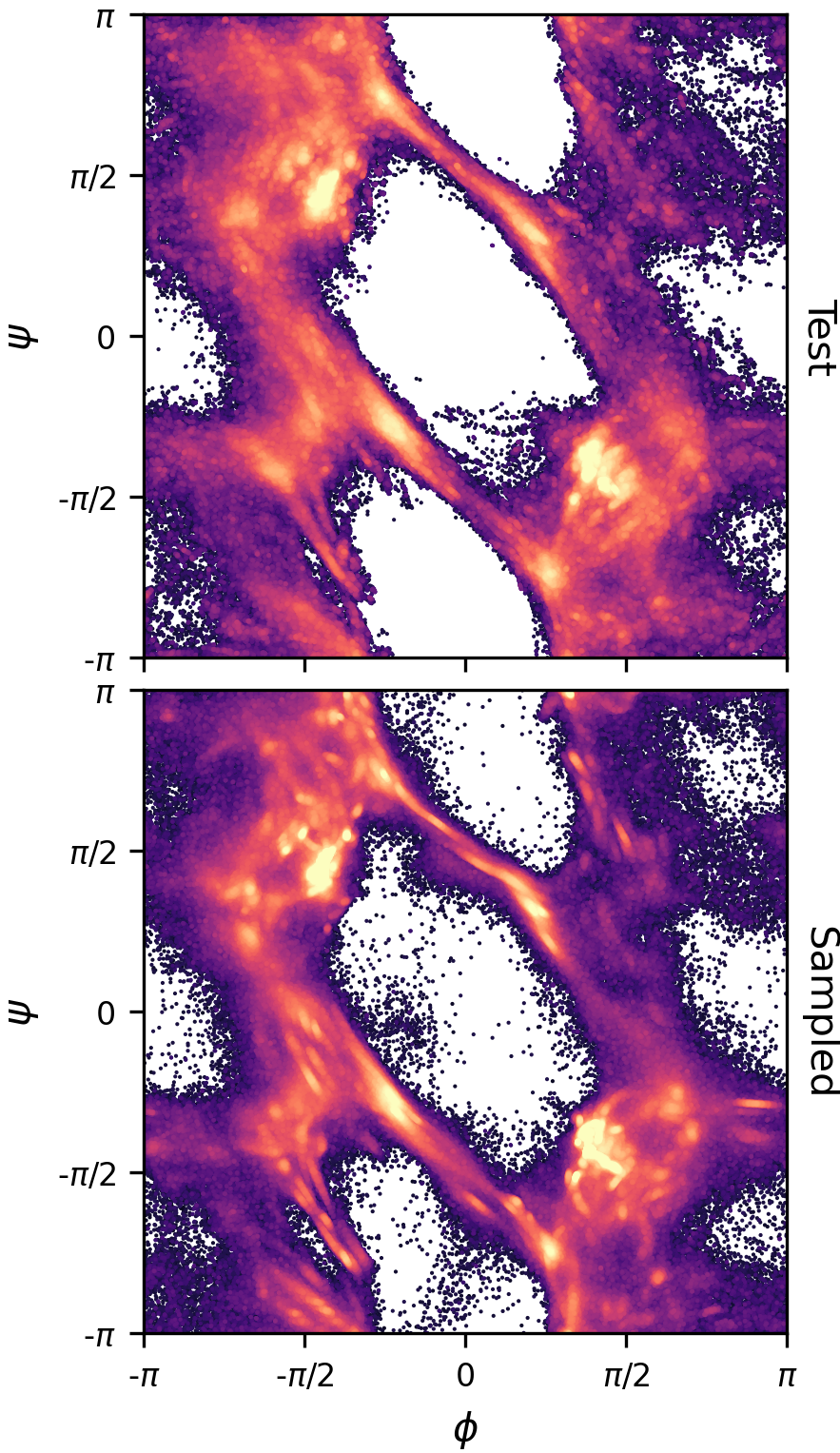}
    \end{minipage}
    \vskip -0.05in
    \caption{Unconditional conformer generation for a \textbf{backbone-only model}. Comparison of the bond angle and dihedral distributions from the held-out test set (orange) and in the \emph{unconditionally} generated redundant internal coordinate samples (blue). The three top left plots correspond to the three bond angle types in each amino acid residue, the three bottom left plots show the three dihedral angles for each residue, and the right shows Ramachandran plots (colored logarithmically by density with high density regions shown in lighter colors). KL divergence is calculated as $D_{\text{KL}}\infdivx{\text{test}}{\text{sampled}}$.}
    \label{fig:angles}
\end{figure}

To understand whether this approach can learn the highly-coupled and underlying distribution of macrocycle conformations, we first trained \shortmethodname{} on macrocycle backbones in the absence of any residue or side-chain features. From a design perspective, diverse backbone sampling alone can help drive inverse peptide design, where specific backbone geometries suggest important sequences. Figure~\ref{fig:angles} (and per-residue distributions in Supplementary Section~\ref{app:ucond}) demonstrates how \shortmethodname{} accurately replicates both angles and dihedrals with tight fidelity across all residue atoms. Furthermore, we generated Ramachandran plots~\cite{Ramachandran1968-mn} alongside our withheld test set to visualize the conditional dependencies between residue torsions. Notably, \shortmethodname{} recapitulates the critical modes of the distribution.

\subsection*{Conditional Generation of Macrocycle Conformational Ensembles}

We subsequently focused on the challenge of conditional generation to understand whether \shortmethodname{} could effectively capture the complex steric and intramolecular effects that dictate macrocycle conformation. Whereas the unconditional model disregards side chains, we now condition generation on molecular features corresponding to each atom in the ring, including side-chain information, stereochemistry, and \textit{N}-methylation (see Supplementary Section~\ref{app:model}) to unambiguously predict all atomic internal coordinates. We evaluate \shortmethodname{} both in the context of all-atom geometries (RMSD), but also evaluate on backbone-only ring geometries (rRMSD, rTFD) as they are critical for macrocycle design~\cite{Bhardwaj2022-jy}. 

\begin{figure}[tb]
    \centering
    \includegraphics[width=\linewidth]{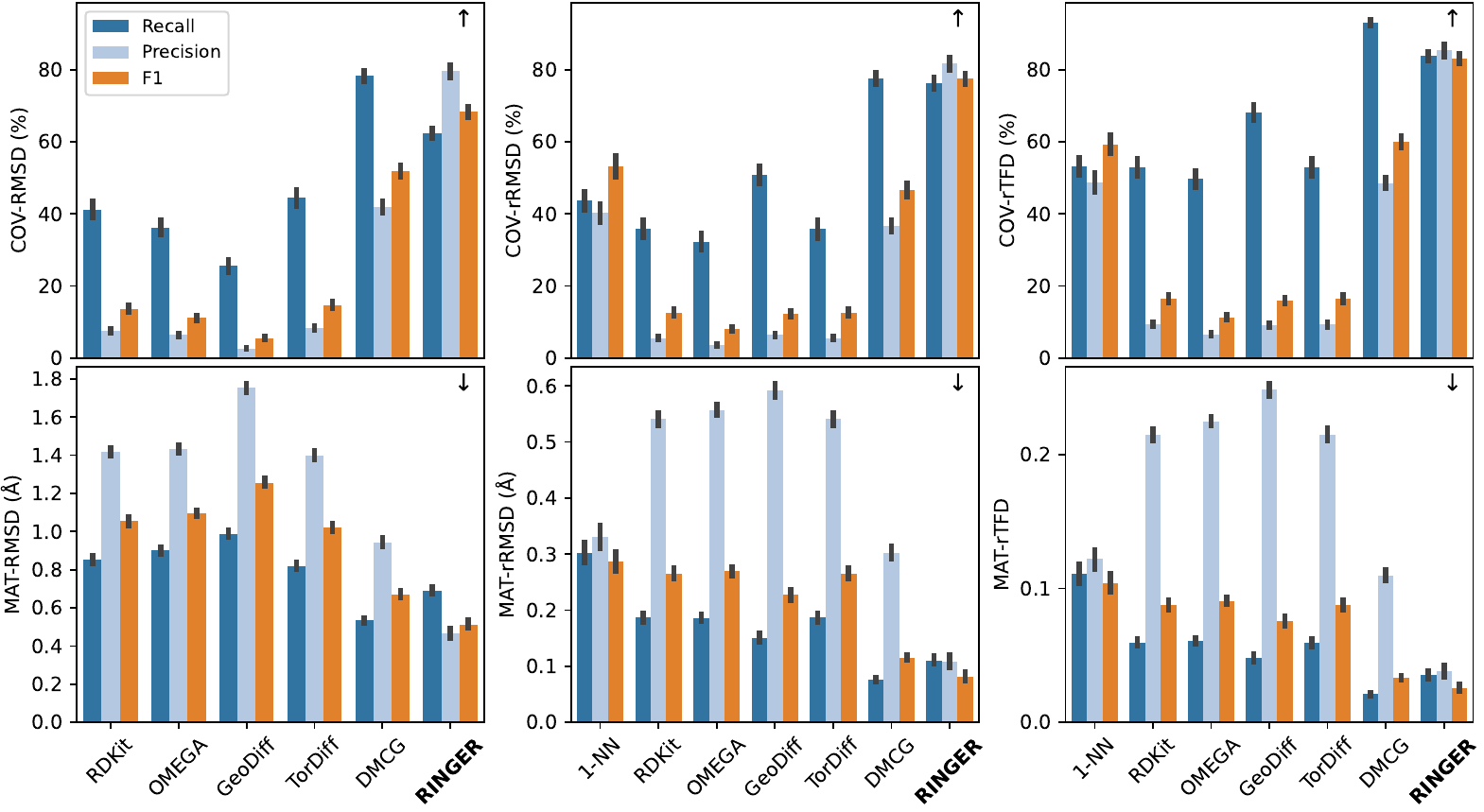}
    \vskip -0.05in
    \caption{\emph{Mean} performance metrics for sequence-conditioned generation of macrocycles. Coverage (COV) is evaluated at a threshold of \SI{0.75}{\angstrom} for all-atom RMSD, \SI{0.1}{\angstrom} for ring-only RMSD (rRMSD), and 0.05 for ring-only TFD (rTFD). For COV, higher scores indicate better model performance ($\uparrow$), and for MAT, lower values are better ($\downarrow$). \emph{All test data} conformers are used for evaluation. Error bars denote 95\% confidence intervals. The exact values are shown in Supplementary Table~\ref{tab:metrics_all_atom}.}
    \label{fig:barplots}
\end{figure}

Comparison of \shortmethodname{} RMSD, rRMSD, and rTFD ensemble metrics against the baselines are shown in Figure~\ref{fig:barplots} and Supplementary Table~\ref{tab:metrics_all_atom}. Here, recall quantifies the proportion of ground truth conformers that are recovered by the model, precision quantifies the quality of the generated ensemble, and F1 is the harmonic mean of recall and precision. We found that RDKit ETKDGv3 and OMEGA Macrocycle mode, both based on distance-geometry approaches, performed similarly across metrics and achieved moderate recall with limited precision. Comparing to other deep learning approaches, we found that both GeoDiff and DMCG offer improved recall over heuristic baselines, with only DMCG generating improved precision and moderate F1-scores. Moreover, the nearest-neighbor baseline, which uses the backbone ensemble from the most similar training molecule (as measured by 2D similarity), provides only modest performance. This baseline demonstrates how side chain identity dictates 3D ensemble geometry, and that deep models cannot trivially memorize backbone geometries in this task. In contrast, \shortmethodname{} achieves excellent performance across all metrics, providing strong recall and excellent precision over all test molecules, with the best F1-score. As demonstrated through rRMSD and rTFD metrics, \shortmethodname{} samples both diverse and high-quality backbone conformations. 

\subsection*{Evaluating Conformer Quality with Post Hoc Optimization}
The CREMP dataset~\cite{grambow023cremp} used in this study leverages the semi-empirical extended tight binding method GFN2-xTB~\cite{Bannwarth2019-rm} for geometry optimization. To enable a level comparison of sampling quality only, we reoptimized all generated samples from all methods with GFN2-xTB (Supplementary Table~\ref{tab:metrics_xtb}, Supplementary Section~\ref{app:xtb_opt}). Notably, we observed a dramatic improvement for xTB-optimized RDKit, OMEGA, GeoDiff, and TorDiff molecules, but only slight improvement in precision. In contrast, xTB-reoptimized samples from \shortmethodname{} see only a minimal but consistent boost in performance, yet still outperform all other approaches. Further analysis of these xTB geometries revealed that \shortmethodname{}-generated samples require smaller conformational adjustments to reach local minima (Supplementary Figure~\ref{fig:rmsd_xtb_changes}, Supplementary Section~\ref{app:xtb_opt}), suggesting that our model can accurately learn the complex distributions directly from the training data. Overall, these results highlight two key strengths of our machine learning approach: 1) careful featurization enables generating high-quality conformations that closely match quantum chemical geometries, and 2) the diffusion scheme provides good sampling diversity.

\subsection*{Structural Analysis of Generated Macrocycles}

\begin{figure}[tb]
    \centering
    \includegraphics[width=\linewidth]{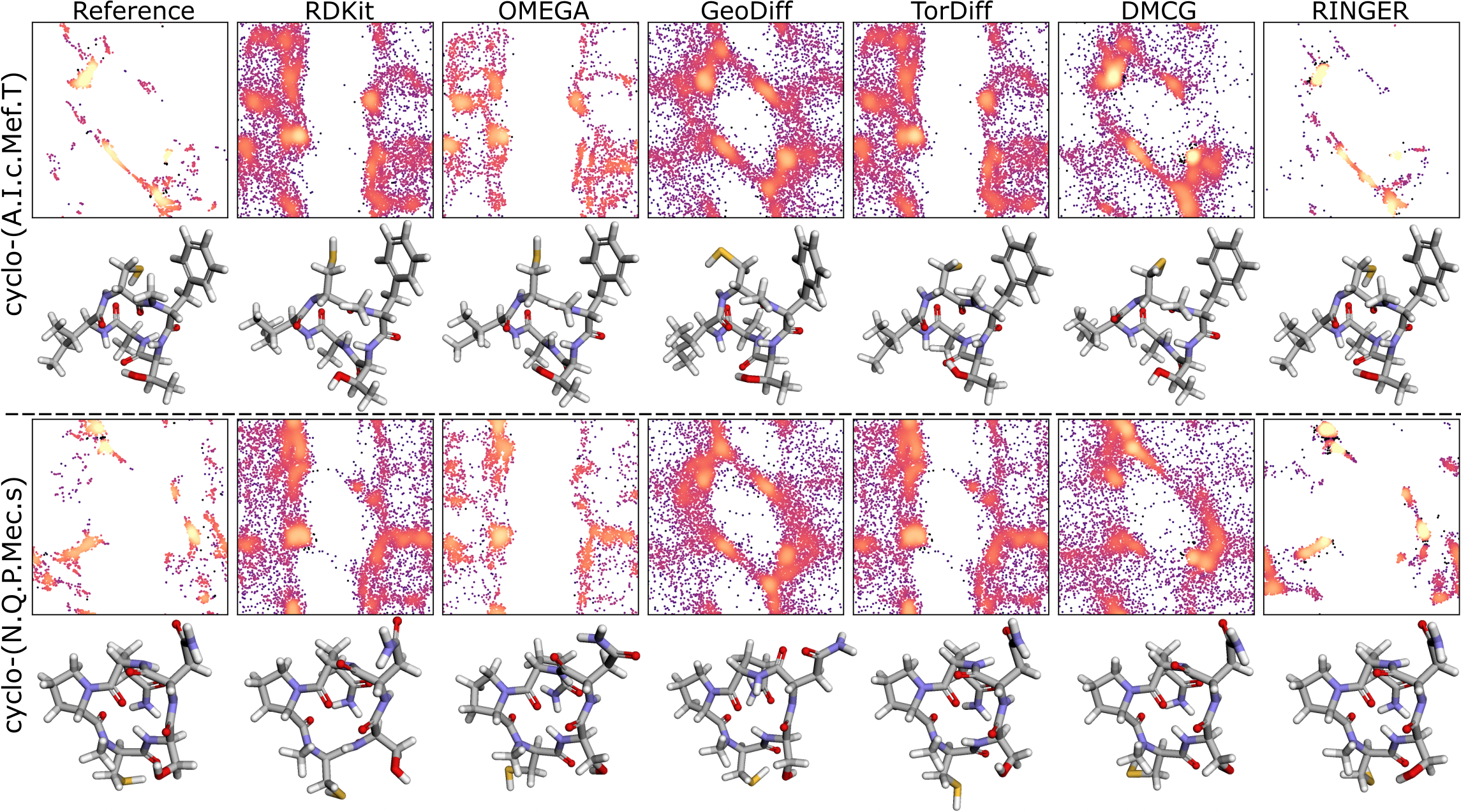}
    \vskip -0.05in
    \caption{Comparison of macrocycle conformational ensembles. \shortmethodname{} accurately generates ensembles as illustrated by Ramachandran plots for individual macrocycle ensembles, whereas other methods generate unrealistic backbone geometries. The individual 3D conformers illustrate the lowest-energy reference structure and the closest matching conformer (based on RMSD) from each method.}
    \label{fig:ramas}
\end{figure}

Although RMSD and TFD evaluate performance quantitatively, analyzing individual ensembles elucidates the qualitative differences in conformer generation processes. Notably, the two distinct macrocycles in Figure~\ref{fig:ramas} possess ensembles characterized by distinctly unique Ramachandran plots. Importantly, most ground truth conformer ensembles exhibit relatively tight distributions characterized by a distinctive set of $\phi$ and $\psi$ angles (Reference, far left). Although all methods can identify relevant low-energy conformers, the overall sampling process generates unrealistic distributions for RDKit, OMEGA, and GeoDiff (note that TorDiff relies on RDKit for backbone sampling, and hence produces an identical plot). In contrast, \shortmethodname{} recapitulates not only the ground state geometry with excellent accuracy, but better captures the entire ensemble distribution. These results demonstrate how \shortmethodname{} achieves strong performance by providing high sample quality and diversity.

\section*{Limitations and Future Directions}
Our studies demonstrate the potential for diffusion-based models to overcome limitations in constrained macrocycle generation, but they are not without drawbacks. Our current work has focused on the CREMP dataset, which is limited to homodetic, 4-, 5-, and 6-mer macrocycles with canonical side chains (albeit including both L and D enantiomers as well as their \textit{N}-methylated versions). Extension to macrocycles with larger ring sizes, non-canonical side chains, and other complex topologies would improve the generalizability of this work, as well as training on datasets generated with higher levels of theory and other methods of treating solvation effects. Additionally, although we demonstrate the effectiveness of a standard, discrete-time diffusion process, our approach is not physically constrained to satisfy macrocyclic geometries and currently requires a post-optimization step. Developing and applying physics-informed diffusion processes with manifold constraints could improve the efficiency of training and sampling of relevant macrocycle backbones.

\section*{Conclusions}
We present \shortmethodname, a new approach for generating macrocycle conformer ensembles that significantly improves sample quality, diversity, and inference. By leveraging strengths of diffusion-based models, we demonstrate how a transformer-based architecture with a cyclic positional encoding results in significant gains over Cartesian-based equivariant models and widely-used distance geometry-based algorithms for both unconditional and conditional structure generation. The present work paves the way for more efficient and accurate computational exploration of conformational space. We anticipate that this approach will more broadly enable rational macrocycle discovery through further development.

\section*{Methods}

\subsection*{Experimental Setup}

\paragraph{Dataset} \label{dataset}
We train and evaluate our approach on the recently published CREMP dataset~\cite{grambow023cremp} that contains 36k homodetic macrocyclic peptides across varying ring sizes (4-mers, 5-mers, and 6-mers corresponding to 12-, 15-, and 18-membered backbone rings), side chains, amino-acid stereochemistry, and \textit{N}-methylation. Each macrocycle in CREMP contains a conformational ensemble sampled with CREST~\cite{Pracht2020-ha,CREST2024}, a multi-start metadynamics algorithm with genetic crossing built on the semi-empirical tight-binding method GFN2-xTB \cite{Bannwarth2019-rm}. We perform stratified random splitting on the data, with a training and validation set of 35,198 molecules (948,158 conformers using a maximum of 30 conformers per molecule), which we split into 90\% training and 10\% validation, and a final test set of 1,000 molecules corresponding to 877,898 distinct conformers (using \emph{all} conformers per molecule within the \SI{6}{kcal/mol} energy threshold defined by CREST). Additional dataset statistics are shown in Supplementary Section~\ref{app:dataset}.

\paragraph{Training \& Sampling} All training is performed on the set of 35k peptides described above, using the 30 lowest-energy conformers per peptide. We train each model on a single NVIDIA A100 GPU for up to 1000 epochs until convergence using the Adam optimizer with 10 warmup epochs. Following work in small-molecule conformer generation~\cite{ganea2021geomol, xu2022geodiff, Jing2022-torsionaldiffusion}, we sample $2K$ conformers for a macrocycle ensemble of $K$ ground-truth conformers (median $K$ = 656) and assess them based on the evaluation criteria below. For full training and sampling details see Supplementary Sections \ref{app:training} and \ref{app:sampling}. 

\paragraph{Evaluation} For unconditional generation, we use the Kullback-Leibler divergence to measure the difference in sample quality. For conditional generation, we evaluate the quality of our generated macrocycles using root-mean-squared-deviation (RMSD) between heavy-atom coordinates, similar to previous work on small-molecule conformer generation. We use several metrics including \textbf{Mat}ching and \textbf{Cov}erage~\cite{xu2021learning, ganea2021geomol, Jing2022-torsionaldiffusion}, and for each we report recall, precision, and F1-score. We also report rRMSD, which is only evaluated on ring atoms. We note that although RMSD is widely used to assess conformer quality, its utility for comparing backbones is more limited, as sampled backbones with highly unrealistic or energetically unfavorable torsions can exhibit low rRMSD values. Therefore, we additionally report the ring torsion fingerprint deviation (rTFD)~\cite{Schulz-Gasch2012-eo, Wang2020-ai} to evaluate the quality of the torsional profiles. RMSD and rRMSD provide a measure of distance between two conformers based on a least-squares alignment of their respective atomic positions, while rTFD gives a normalized measure of matched torsion angles between backbone geometries. Supplementary Section~\ref{app:evaluation} defines the evaluation metrics in detail.

\paragraph{Baselines} We provide benchmarks of our method against open-source and commercial toolkits RDKit ETKDGv3 (for macrocycles)~\cite{Wang2020-ai}, OMEGA Macrocycle Mode~\cite{openeyeomega-42}, GeoDiff~\cite{xu2022geodiff}, DMCG~\cite{zhu2022dmcg}, and Torsional Diffusion~\cite{Jing2022-torsionaldiffusion} (TorDiff). We retrained GeoDiff, DMCG, and Torsional Diffusion on CREMP, as the base models trained on small molecules provided poor performance. To better understand ring diversity, we additionally included a nearest neighbor model (1-NN) based on maximum graph similarity (2D) for ring-evaluation metrics only.

\bibliography{main}

\section*{Acknowledgements}

We thank Ben Sellers and Christian Cunningham for insightful discussions on macrocycles and peptide therapeutics. We also thank members of the Departments of Peptide Therapeutics and Discovery Chemistry for helpful feedback and discussions.

\section*{Author Contributions Statement}

C.A.G. and K.V.C. conceived the study and implemented the code. C.A.G, H.W., and K.V.C. designed the experiments and performed modeling and analysis. N.L.D., G.S., and T.B. provided detailed guidance on diffusion modeling and optimization. All authors contributed to writing and editing and approved the final manuscript.

\section*{Code and Data Availability}
All code for training, sampling, and evaluation in this study is available at \href{https://github.com/genentech/ringer}{github.com/Genentech/RINGER}. We include the exact training and test data splits as well as the trained model. The CREMP dataset~\cite{grambow023cremp} is available for download from \href{https://zenodo.org/doi/10.5281/zenodo.7931444}{zenodo.org/doi/10.5281/zenodo.7931444}. 

\section*{Competing Interests}
This research is sponsored by Genentech, Inc. All authors are employees of Genentech, Inc. and shareholders of Roche.

\newpage

\appendix
\renewcommand\thesection{S.\arabic{section}}
\renewcommand\thesubsection{S.\arabic{section}.\arabic{subsection}}

\renewcommand{\thefigure}{S.\arabic{figure}}
\setcounter{figure}{0}
\renewcommand{\thetable}{S.\arabic{table}}
\setcounter{table}{0}
\renewcommand{\theequation}{S.\arabic{equation}}
\setcounter{equation}{0}

\section*{Supplementary Information for Accurate and Efficient Structural Ensemble Generation of Macrocyclic Peptides using Internal Coordinate Diffusion}

\section{Glossary}
\label{app:glossary}

\begin{table}[H]
\caption{Glossary of notation and terms used in the methods section.}
\vspace{-10pt}
\label{table:glossary}
\centering
\begin{tabular}{ll}
\toprule
\textbf{Symbol} & \textbf{Description} \\
\midrule
\multicolumn{2}{c}{\textbf{Molecular Representation and Coordinates}} \\
\midrule
$\mathcal{G}$ & Macrocycle graph, where $\mathcal{G} = (\mathcal{V}, \mathcal{E})$. \\
$\mathcal{V}$ & Set of atom vertices in a macrocycle graph $\mathcal{G}$. \\
$\mathcal{E}$ & Set of edges (bonds) between the atoms in a macrocycle graph $\mathcal{G}$. \\
$\mathcal{C}$ & Set (ensemble) of conformers for a macrocycle, where $\mathcal{C} = \{c_1, c_2, \dots, c_K\}$. \\
$\mathcal{D}$ & Bond distances in a conformer ensemble $\mathcal{C}$. \\
$\Theta$ & Bond angles in a conformer ensemble $\mathcal{C}$. \\
$\mathcal{T}$ & Dihedral (torsional) angles in a conformer ensemble $\mathcal{C}$. \\
$\bm{\upxi}$ & Vector of all ring Cartesian coordinates in a conformer. \\
$\mathbf{d}$ & Vector of all ring bond distances in a conformer. \\
$\bm{\uptheta}$ & Vector of all ring bond angles in a conformer. \\
$\bm{\uptau}$ & Vector of all ring dihedral (torsional) angles in a conformer. \\
$d_{i,j}$ & Bond distance between atoms $v_i$ and $v_j$. \\
$\theta_{i,j,k}$ & Bond angle between atoms $v_i$, $v_j$, and $v_k$. \\
$\tau_{i,j,k,l}$ & Dihedral (torsional) angle between atoms $v_i$, $v_j$, $v_k$, and $v_l$. \\
$\phi$ & Dihedral angle of bond between nitrogen and $\alpha$-carbon. \\
$\psi$ & Dihedral angle of bond between $\alpha$-carbon and carbonyl-carbon. \\
$\omega$ & Dihedral angle of bond between carbonyl-carbon and nitrogen (peptide bond). \\
\midrule
\multicolumn{2}{c}{\textbf{Encoder Model}} \\
\midrule
$\mathbf{a}_i$ & Atom features for ring vertex $v_i$. \\
$\bm{\uptheta}_i$ & Bond angles corresponding to ring vertex $v_i$ (1 ring \& 5 side chain). \\
$\bm{\uptau}_i$ & Dihedral (torsional) angles corresponding to ring vertex $v_i$ (1 ring \& 5 side chain). \\
$\mathbf{x}_i$ & Internal coordinates corresponding to vertex $v_i$, where $\mathbf{x}_i=\bm{\uptheta}_i \oplus \bm{\uptau}_i$. \\
$\mathbf{v}_i$ & Input/hidden representation for ring vertex $v_i$. \\
$\mathbf{z}_i$ & Self-attention output for vertex $v_i$. \\
$\alpha_{ij}$ & Attention probability between vertices $v_i$ and $v_j$. \\
$e_{ij}$ & Unnormalized attention score between vertices $v_i$ and $v_j$. \\
$d_z$ & Attention head dimensionality. \\
$\mathbf{p}_{ij}^K$ & Cyclic relative positional embedding between vertices $v_i$ and $v_j$. \\
$\mathbf{W}^K$ & Key matrix. \\
$\mathbf{W}^Q$ & Query matrix. \\
$\mathbf{W}^V$ & Value matrix. \\
$\mathbf{W}^D$ & Graph-distance embedding matrix. \\
\midrule
\multicolumn{2}{c}{\textbf{Diffusion Process}} \\
\midrule
$\mathbf{x}_t$ & Noised internal coordinates (bond angles and torsions) at time step $t$. \\
$q\giventhat{\mathbf{x}_t}{\mathbf{x}_{t-1}}$ & Forward transition probability. \\
$p_{\Xi}\giventhat{\mathbf{x}_{t-1}}{\mathbf{x}_t}$ & Diffusion model (reverse transition probability) parameterized by $\Xi$. \\
$\beta_t$ & Variance from cosine variance schedule at time step $t$. \\
$\bm{\upepsilon}_t$ & Noise scale at time step $t$. \\
\midrule
\multicolumn{2}{c}{\textbf{Miscellaneous}} \\
\midrule
$\hat{\cdot}$ & Denotes predicted/generated quantity. \\
$w(\cdot)$ & Function to wrap within $[-\pi,\pi)$ range, $w(\tau) \coloneqq (\tau+\pi) \bmod (2\pi) - \pi$. \\
$\delta$ & Threshold for evaluating Coverage metric. \\
$\oplus$ & Concatenation. \\
\bottomrule
\end{tabular}%
\end{table}%

\newpage
\section{Dataset Description}
\label{app:dataset}

We leverage the recently described Conformer-Rotamer Ensembles of Macrocyclic Peptides (CREMP) dataset~\cite{grambow023cremp} that contains 36,198 unique macrocyclic peptide sequences and their corresponding ensembles, totaling 31.3 million conformers. All conformers in CREMP were optimized using the GFN2-xTB semi-empirical quantum chemistry method~\cite{Bannwarth2019-rm}. GFN2-xTB incorporates physics-based terms for dispersion, electrostatics, hydrogen bonding, and other quantum mechanical effects into a self-consistent tight-binding framework. This provides a balance of reasonable accuracy and computational efficiency, bridging the gap between fast but inaccurate force fields and high-level yet expensive DFT methods.

The conformational sampling was performed using CREST~\cite{Pracht2020-ha,CREST2024}, which combines metadynamics enhanced sampling with GFN2-xTB for energies and forces. Metadynamics iteratively adds biasing potentials to guide sampling to unexplored areas of conformational space, enabling more thorough sampling than conventional molecular dynamics. By pairing metadynamics with GFN2-xTB, CREST balances accuracy and computational efficiency when generating macrocycle ensembles. However, metadynamics remains expensive, requiring thousands of CPU hours per molecule. The CREMP dataset hence provides extensive high-quality training data of macrocycle conformers. Our work builds on CREMP, by developing deep generative models to approximate these computationally demanding physics-based methods for sampling.

\begin{table}[H]
\centering
\caption{Dataset statistics for CREMP~\cite{grambow023cremp}.}
\label{tab:stats}
\begin{tabular}{cccccccc}
\toprule
\textbf{Residues} & \textbf{Molecules} & \multicolumn{6}{c}{\textbf{Conformers}} \\
& & Count & Mean & Median & Std.\ Dev.\ & Min.\ & Max.\ \\
\midrule
4 & 17,842 & 12,205,128 & 684 & 508 & 677 & 1 & 12,268 \\
5 & 13,644 & 14,134,609 & 1,036 & 825 & 824 & 6 & 8,486 \\
6 & 4,712 & 4,921,068 & 1,044 & 879 & 764 & 28 & 5,619 \\
\midrule
Total & 36,198 & 31,260,805 & 864 & 656 & 768 & 1 & 12,268 \\
\bottomrule
\end{tabular}%
\end{table}%

\section{Training Details}
\label{app:training}

We adapt a discrete-time diffusion scheme that formulates the forward transition probability using a wrapped normal distribution,

\begin{equation}
\begin{split}
q\giventhat*{\mathbf{x}_t}{\mathbf{x}_{t-1}} &= \mathcal{N}_{\text {wrapped}}\left(\mathbf{x}_t ; \sqrt{1-\beta_t} \mathbf{x}_{t-1}, \beta_t \mathbf{I}\right) \\
&= \frac{1}{\beta_t \sqrt{2\pi}} \sum_{\mathbf{k} \in \mathbb{Z}^n} \exp \left(\frac{-\left\|\mathbf{x}_t-\sqrt{1-\beta_t} \mathbf{x}_{t-1}+2 \pi \mathbf{k}\right\|^2}{2 \beta_t^2}\right)
\end{split}
\end{equation}

\noindent instead of a standard normal distribution~\cite{wu2022protein, Jing2022-torsionaldiffusion}, where $\mathbf{x}_t$ represents the noised internal coordinates (bond angles and torsions) at time step $t$. The diffusion model, $p_{\Xi}\giventhat{\mathbf{x}_{t-1}}{\mathbf{x}_t}$, parameterized by $\Xi$, reverses the process to denoise a wrapped normal distribution toward the data distribution. In the conditional setting, we further guide the diffusion process by learning $p_{\Xi}\giventhat*{\mathbf{x}_{t-1}}{\mathbf{x}_t, \mathcal{G}}$ in order to draw samples from the ensemble for a specific macrocycle, $\mathcal{G}$. We use the same cosine variance schedule as Wu et al.~\cite{wu2022protein} and Nichol \& Dhariwal~\cite{pmlr-v139-nichol21a} for $\beta_t \in (0, 1)_{t=1}^T$, but with significantly fewer time steps ($T=20$). $p_{\Xi}\giventhat{\mathbf{x}_{t-1}}{\mathbf{x}_t}$ and $p_{\Xi}\giventhat*{\mathbf{x}_{t-1}}{\mathbf{x}_t, \mathcal{G}}$ are trained using the simplified objective from Ho et al.~\cite{Ho-denoising-2020} to train a neural network, $\bm{\upepsilon}_{\Xi}(\mathbf{x}_t, t)$, to predict the noise present at a given time step by minimizing a smooth L1 loss \cite{Girshick2015} wrapped by $w(\mathbf{x}) = (\mathbf{x}+\pi) \bmod (2\pi) - \pi$:

\begin{equation}
\begin{split}
   \mathbf{d}_w &= w\left(\bm{\upepsilon} - \bm{\upepsilon}_{\Xi} \left(w \left(\sqrt{\bar{\alpha}_t}\mathbf{x}_0 + \sqrt{1-\bar{\alpha}_t}\bm{\upepsilon}\right), t\right)\right) \\
L_w &= \frac{1}{N} \sum_{i=1}^N
\begin{cases}
    0.5 \frac{d_{w,i}^2}{\beta_L} & \text{if } |d_{w,i}| < \beta_L\\
    |d_{w,i}| - 0.5\beta_L       & \text{otherwise}
\end{cases}
\end{split}
\end{equation}

\noindent with $\beta_L=0.1\pi$ as the transition point between L1 and L2 regimes \cite{wu2022protein}, $\alpha_t = 1-\beta_t$, and $\bar{\alpha}_t = \prod_{s=1}^t \alpha_s$. We sample time steps uniformly from $t \sim U(0,T)$ during training and shift the bond angles and dihedrals using the element-wise means from the training data.

\section{Sampling Details}
\label{app:sampling}

During inference, we first sample $\mathbf{x}_T$ from a wrapped normal distribution and iteratively generate $\mathbf{x}_0$ from $t=T$ to $t=1$ using

\begin{equation}
    \mathbf{x}_{t-1} = w\left( \frac{1}{\sqrt{\alpha_t}} \left( \mathbf{x}_t - \frac{1-\alpha_t}{\sqrt{1-\bar{\alpha}_t}} \bm{\upepsilon}_{\Xi} (\mathbf{x}_t,t) \right) + \sigma_t \mathbf{x} \right)
\end{equation}

where $\sigma_t = \sqrt{\beta_t (1-\bar{\alpha}_{t-1})/(1-\bar{\alpha}_t)}$ is the variance of the reverse process and $\mathbf{z} = \mathcal{N}(\mathbf{0},\mathbf{I})$ if $t>1$ and $\mathbf{z}=\mathbf{0}$ otherwise~\cite{wu2022protein}.

\section{Model Details and Hyperparameters}
\label{app:model}

Our model is a BERT transformer~\cite{devlin-etal-2019-bert} with graph-based, cyclic relative positional encodings described in Equations~\eqref{eq:cyclic1} and \eqref{eq:cyclic2}. The model input is a sequence of internal coordinates (and atom features for the conditional model). We linearly upscale the model input (bond angles and dihedrals) and separately upscale the atom features. Angles and atom features are then concatenated. The time step is embedded using random Fourier embeddings~\cite{song2021scorebased} and added to the upscaled input. The combined embeddings are passed through the BERT transformer, the output of which is passed through a two-layer feed-forward network with GELU activation and layer normalization. Relevant hyperparameters are shown in Table~\ref{table:hyperparams}.

\begin{table}[H]
\caption{Hyperparameters.}
\label{table:hyperparams}
\centering
\begin{tabular}{ll}
\toprule
\textbf{Parameter} & \textbf{Value} \\
\midrule
Angle embedding size & 256 \\
Atom feature embedding size & 128 \\
Encoder layer dimensionality (hidden size) & 384 \\
Number of hidden layers & 12 \\
Number of attention heads & 12 \\
Feed-forward layer dimensionality (intermediate size) & 512 \\
\midrule
Optimizer & AdamW \\
Learning rate & \num{e-3} (restarted with \num{5e-4}) \\
Maximum number of epochs & 1000 \\
Warmup epochs & 10 \\
Batch size & 8192 \\
\bottomrule
\end{tabular}%
\end{table}%

To condition on the atom sequence, we encode each atom using features of the atom itself and a Morgan fingerprint representation of the side chain attached to the atom (including the atom itself). The atom features include the atomic number, a chiral tag (L, D, or no chirality), aromaticity, hybridization, degree, valence, number of hydrogens, charge, sizes of rings that the atom is in, and the number of rings that the atom is in. The Morgan fingerprint is a count fingerprint with radius 3 and size 32.

\section{Optimization for Back Conversion to Cartesian Ring Coordinates}
\label{app:opt}

\begin{figure}[H]
    \centering
    \includegraphics[width=0.6\linewidth]{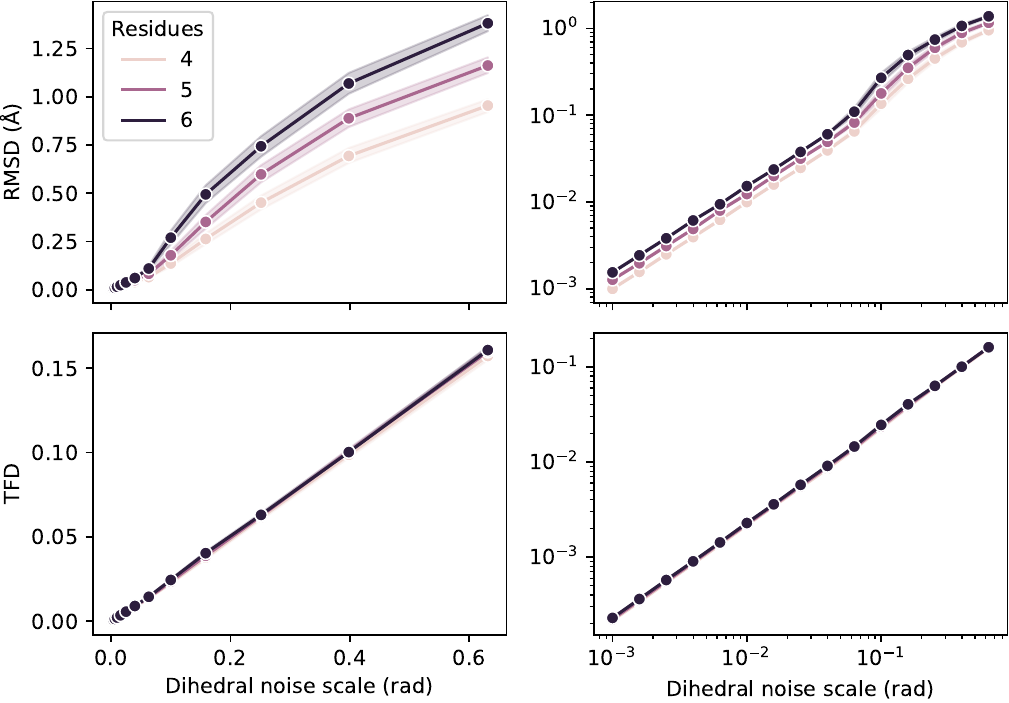}
    \vskip -0.05in
    \caption{Our constrained optimization procedure is robust to noise as illustrated by a synthetic test of applying noise to the dihedral angles, recovering Cartesian coordinates using Equation~\eqref{eq:opt}, and comparing to the initial geometry in terms of rRMSD and rTFD.}
    \label{fig:optimization}
\end{figure}

To convert from the set of redundant internal coordinates predicted by the model back to Cartesian coordinates, we solve the optimization in Equation~\eqref{eq:opt} to obtain a set of Cartesian coordinates that exactly satisfies the known bond distances in the ring. To demonstrate that this procedure is robust to noise, we repeatedly embed 4-, 5-, and 6-mer backbones in 3D using RDKit distance geometry, extract their (redundant) internal coordinates, and add noise to the dihedral angles at different noise scales (standard deviation of a normal distribution) while ensuring that angles always remain in the $[-\pi,\pi)$ range. This creates a set of inconsistent, redundant dihedral angles, i.e., there exists no direct correspondence in Cartesian coordinates. We recover a possible Cartesian configuration using Equation~\eqref{eq:opt} and compute rRMSD and rTFD for the ring atoms compared to the ``true'' internal coordinates from the RDKit geometry. Figure~\ref{fig:optimization} shows that even moderate errors ($\sim$\SI{0.1}{rad}) result in very small errors in terms of both rRMSD ($\sim$\SI{0.1}{\angstrom}) and rTFD ($\sim$0.02).

Notably, the optimization problem in Equation~\eqref{eq:opt} is non-convex and requires a suitable initial guess to perform well. We assign this initial guess by obtaining a Cartesian geometry using the approach of sequentially setting atom positions according to the sequence of bond distances, angles, and torsions starting from each of the atoms in the ring. We then average the so-generated $n_{\text{ring}}$ sets of Cartesian coordinates and use the resulting coordinates as the initial guess.

\newpage
\section{Overall Method Overview}
\label{app:overview}

\begin{figure}[H]
    \centering
    \includegraphics[width=0.93\linewidth]{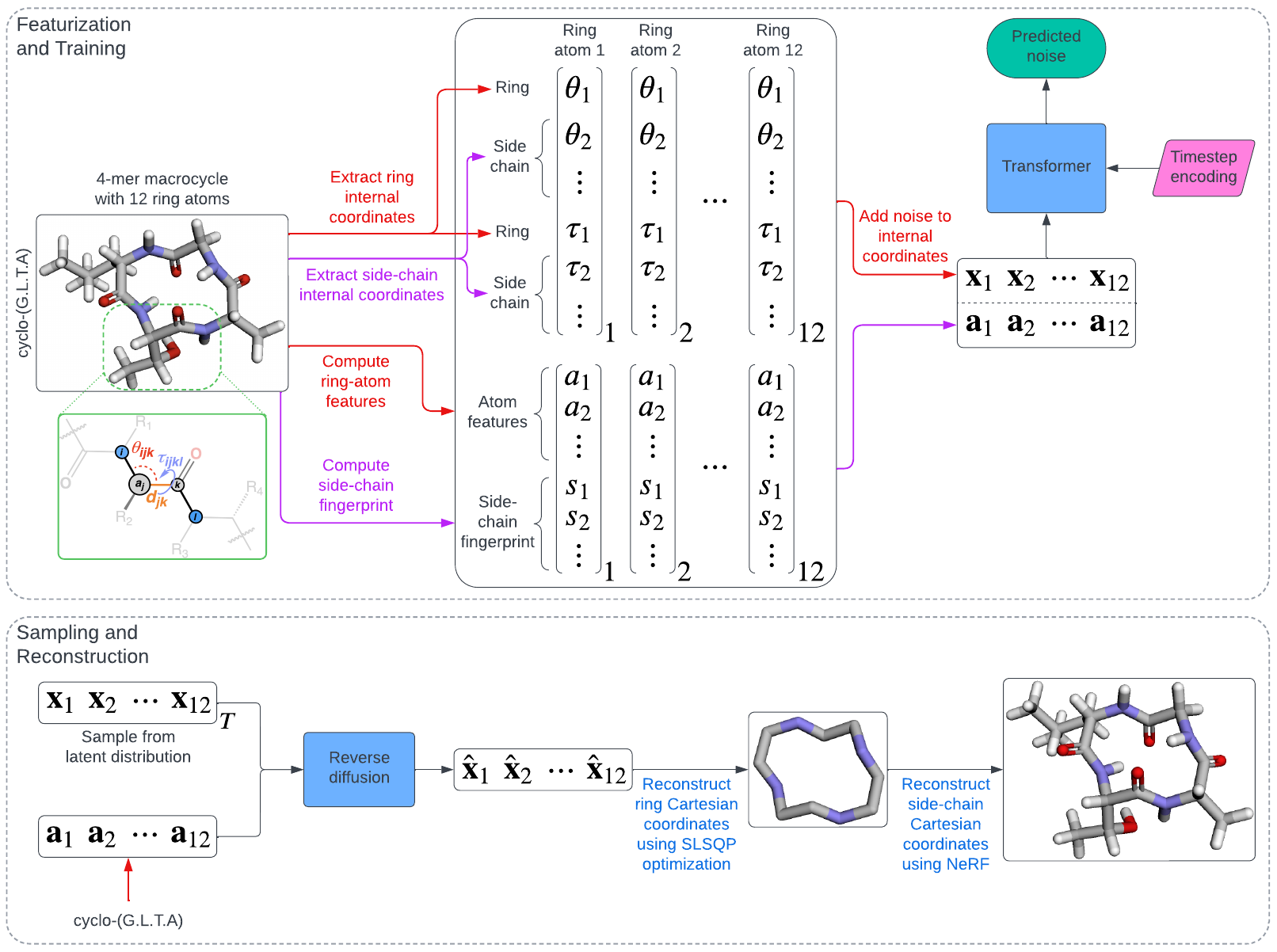}
    \vskip -0.05in
    \caption{Overall training and sampling procedure.}
    \label{fig:flowchart}
\end{figure}%
\begin{figure}[H]
    \centering
    \includegraphics[width=\linewidth]{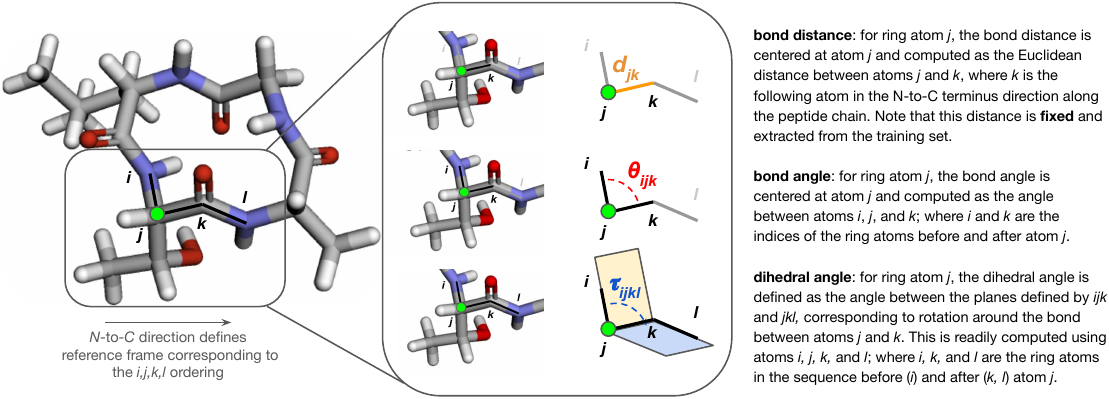}
    \vskip -0.05in
    \caption{\textbf{Schematic representation of how internal coordinates are assigned to each ring atom.} For each atom in the macrocycle, we define a consistent direction based on the N-to-C directionality of the peptide backbone. Neighboring atoms then define the key vectors and planes used to calculate bond distances, angles, and torsions. Specifically, for an atom $j$, the N-to-C direction defines a set of indices ($i$, $j$, $k$, $l$) where $i = j - 1$, $k = j + 1$, and $l = j + 2$ along the macrocycle backbone. Internal coordinates are readily calculated from their Cartesian coordinates through standard geometry calculations.}
    \label{fig:ring_ic_extraction}
\end{figure}

\newpage
\section{Software}
\label{app:software}

All experiments were performed using Python and standard numerical libraries. For cheminformatics analysis, all molecules were processed using either OpenEye Applications and Toolkits~\cite{openeyeomega-42} or the open-source cheminformatics library RDKit~\cite{landrum-rdkit}. We implemented all experiments in Python using
PyTorch~\cite{Paszke_PyTorch_An_Imperative_2019} and PyTorch Lightning~\cite{Falcon_PyTorch_Lightning_2019}. Transformers were implemented using BERT models within HuggingFace Transformers~\cite{wolf-etal-2020-transformers}.

\section{Hardware}
Each model was trained on a single NVIDIA A100 GPU with 80~GB VRAM using 12~CPUs for data loading and 96~GB of memory.

\section{Evaluation}
\label{app:evaluation}

To measure both diversity and quality of the generated ensembles, we follow previous work and leverage four RMSD-based metrics~\cite{xu2021learning,ganea2021geomol}. The \emph{recall}-based \textbf{Cov}erage metric measures the percentage of correctly generated conformers at a certain RMSD threshold, $\delta_{\text{RMSD}}$. For a ground-truth ensemble $\mathcal{C}$ and a generated ensemble $\hat{\mathcal{C}}$:

\begin{equation}
    \text{RMSD-COV-R} ( \hat{\mathcal{C}},\mathcal{C}) = \frac{1}{\lvert \mathcal{C} \rvert} \left\lvert \left\{ c \in \mathcal{C} : \exists \hat{c} \in \hat{\mathcal{C}}, \text{RMSD}(\hat{c},c) \leq \delta_{\text{RMSD}} \right\} \right\rvert
\end{equation}

The \emph{recall}-based \textbf{Mat}ching metric measures the average RMSD across the closest-matching (minimum-RMSD) generated conformer for each ground-truth conformer:

\begin{equation}
    \text{RMSD-MAT-R} ( \hat{\mathcal{C}},\mathcal{C} ) = \frac{1}{\lvert \mathcal{C} \rvert} \sum_{c \in \mathcal{C}} \min_{\hat{c} \in \hat{\mathcal{C}}} \text{RMSD} (\hat{c}, c)
\end{equation}

The other two RMSD-based metrics are \emph{precision} metrics that are defined identically, except that the ground-truth and generated ensembles are switched, and therefore constitute a measure of how many generated conformers are of high quality. Similarly, we compute four RMSD metrics on the ring atoms \emph{only}, indicated using rRMSD.

Analogous to the RMSD-based metrics, we define four metrics based on ring torsion fingerprint deviation (rTFD)~\cite{Schulz-Gasch2012-eo, Wang2020-ai} to measure diversity and quality in terms of the torsional profiles of the generated rings:

\begin{equation}
    \text{rTFD-COV-R} ( \hat{\mathcal{C}},\mathcal{C}) = \frac{1}{\lvert \mathcal{C} \rvert} \left\lvert \left\{ c \in \mathcal{C} : \exists \hat{c} \in \hat{\mathcal{C}}, \text{rTFD}(\hat{c},c) \leq \delta_{\text{rTFD}} \right\} \right\rvert
\end{equation}

\begin{equation}
    \text{rTFD-MAT-R} ( \hat{\mathcal{C}},\mathcal{C} ) = \frac{1}{\lvert \mathcal{C} \rvert} \sum_{c \in \mathcal{C}} \min_{\hat{c} \in \hat{\mathcal{C}}} \text{rTFD} (\hat{c}, c)
\end{equation}

rTFD quantifies how well the macrocycle torsion angles match between two conformers and is given by~\cite{Wang2020-ai}:

\begin{equation}
    \text{rTFD}(\hat{c}, c) = \frac{1}{n_{\text{ring}}} \sum_{i=1}^{n_{\text{ring}}} \frac{1}{\pi} \left\lvert w\left( \tau_i(\hat{c}) - \tau_i(c) \right) \right\rvert
\end{equation}

\noindent where $\tau_i(c)$ extracts the $i$-th macrocycle torsion angle of conformer $c$ and $w(\cdot)$ ensures that the deviation is wrapped correctly around the $[-\pi,\pi)$ boundary. Each torsion deviation is normalized by the maximum (absolute) deviation, $\pi$, so that rTFD lies in $[0,1]$.

For both COV and MAT we also compute an F1 score, which is defined as the harmonic mean between precision and recall.

\section{Conformer Generation Baselines}
\label{app:baselines}

\paragraph{RDKit ETKDGv3} RDKit baselines used ETKDGv3~\cite{Riniker2015-kk, Wang2020-ai} with macrocycle torsion preferences. We first embedded up to $2K$ conformers (where $K$ is the number of true conformers) using \texttt{EmbedMultipleConfs} with random coordinate initialization (\texttt{useRandomCoords=True}), which has been shown to be beneficial for generating macrocycle geometries~\cite{Wang2020-ai}. Conformers were subsequently optimized using MMFF94~\cite{Halgren_undated-kw} as implemented in RDKit and sorted by energy. Finally, the sorted conformers were filtered based on heavy-atom RMSD with a threshold of \SI{0.1}{\angstrom}.

\paragraph{OpenEye OMEGA: Macrocycle Mode} OMEGA baselines were performed using OpenEye Applications (\texttt{2022.1.1}) with OMEGA (\texttt{v.4.2.0})~\cite{Hawkins2010-xq, Hawkins2012-aj} in \texttt{macrocycle} mode~\cite{Spellmeyer1997-wx}. Conformational ensembles were generated with the following \texttt{macrocycle} settings: \texttt{maxconfs=}$2K$, \texttt{ewindow=20}, \texttt{rms=0.1}, \texttt{dielectric\_constant=5.0}, where $K$ corresponds to the number of ground truth conformers from the original CREST ensemble in the CREMP dataset. The dielectric constant was set to 5.0 (chloroform) to most closely mimic the implicit chloroform solvation used in CREMP.

\paragraph{GeoDiff} We used the original paper implementation and code (\href{https://github.com/MinkaiXu/GeoDiff}{github.com/MinkaiXu/GeoDiff}) of GeoDiff from Xu et al.~\cite{xu2022geodiff}, which we retrained to convergence on the CREMP dataset with the same data splits. We used the same experimental details as the GEOM-Drugs model from Xu et al.~\cite{xu2022geodiff}. As with all the other methods, we evaluated GeoDiff by sampling $2K$ conformers for each molecule. Inference for GeoDiff uses 5,000 time steps, which required more than \SI{24}{h} for all test set molecules on 20 A100 GPUs.

\paragraph{Torsional Diffusion} We used the original paper implementation and code (\href{https://github.com/gcorso/torsional-diffusion}{github.com/gcorso/torsional-diffusion}) of Torsional Diffusion (TorDiff) from Jing et al.~\cite{Jing2022-torsionaldiffusion}, which we retrained to convergence on the CREMP dataset with the same data splits. We used the same experimental details as the GEOM-Drugs model from Jing et al.~\cite{Jing2022-torsionaldiffusion} with slight modifications to improve the performance for the macrocycle structure task: Prior to training, a conformer matching procedure is necessary to account for the distributional shift between RDKit local structures and xTB-optimized geometries. In order to account for this, we seeded the training local structures using the ground-truth xTB-optimized conformers with subsequent MMFF94 force-field optimization. At inference time, we seeded the local structures by generating RDKit conformers as described above, which very significantly boosted performance of TorDiff. The original implementation~\cite{Jing2022-torsionaldiffusion} only seeds conformers using default RDKit distance geometry embedding parameters, which generates low-quality conformers for macrocycles and leads to very poor performance.

\paragraph{DMCG} We used the original paper implementation and code (\href{https://github.com/DirectMolecularConfGen/DMCG}{github.com/DirectMolecularConfGen/DMCG}) of DMCG from Zhu et al.~\cite{zhu2022dmcg}, which we retrained to convergence on the CREMP dataset with the same data splits. After a basic hyperparameter search, we used the same experimental details as the GEOM-Drugs model from Zhu et al.~\cite{zhu2022dmcg}. Regardless of trainer hyperparameters, strong overfitting occurred after two cycles through the cyclic learning rate scheduler based on the validation loss. Therefore, we selected the best checkpoint by evaluating the first two validation loss minima using the metrics described in Supplementary Section~\ref{app:evaluation} and selecting the best one. It should be noted that the DMCG model has an \emph{order of magnitude more parameters} than the other methods. As with all the other methods, we evaluated DMCG by sampling $2K$ conformers for each molecule.

\paragraph{1-NN (Nearest Neighbor Baseline)} As an instance-based baseline for evaluating macrocycle backbones only (with rRMSD and rTFD), we use a simple 2D similarity approach to find the nearest sequence neighbor for a test molecule within the training set. Each macrocycle is featurized using a residue-wise RDKit Morgan fingerprint, and we calculate the maximum similarity by exhaustively comparing all possible sequence alignments across each training set macrocycle. The backbone is then extracted from the training set molecules and its conformers are used for ring evaluation as above.

\newpage
\section{Unconditional Backbone Generation: Additional Results and Plots}

As a proof-of-concept, we trained a backbone-only unconditional model to understand whether our approach could accurately model the complex distribution of coupled coordinates. We provide additional plots below, split by macrocycle size, to demonstrate that our model is expressive enough to capture the critical modes of these distributions with granularity.
\label{app:ucond}

\begin{figure}[H]
    \centering
    \includegraphics[width=0.75\linewidth]{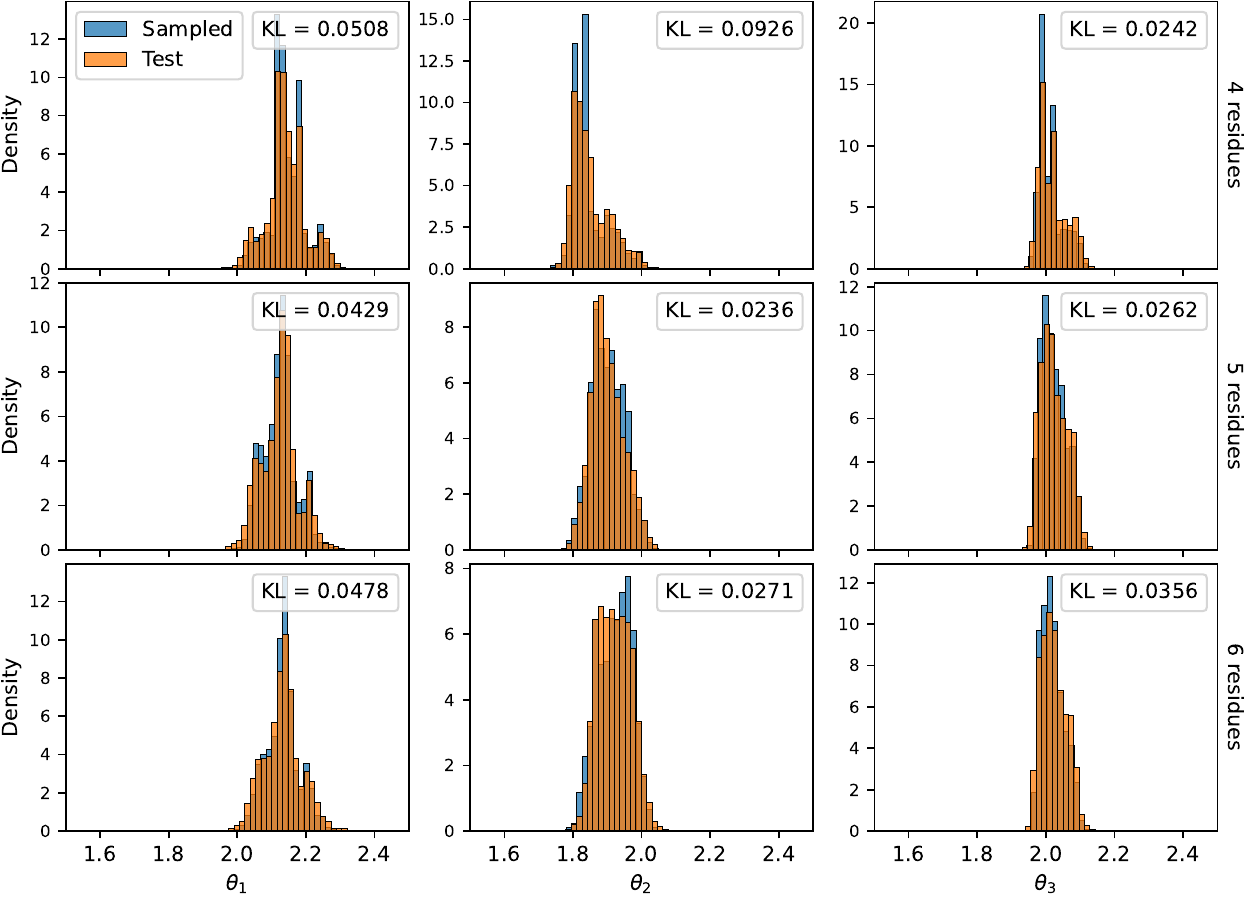}
    \vskip -0.05in
    \caption{Bond angle distributions split by number of residues for the backbone-only unconditional model.}
    \label{fig:angles_residues}
\end{figure}

\begin{figure}[H]
    \centering
    \includegraphics[width=0.75\linewidth]{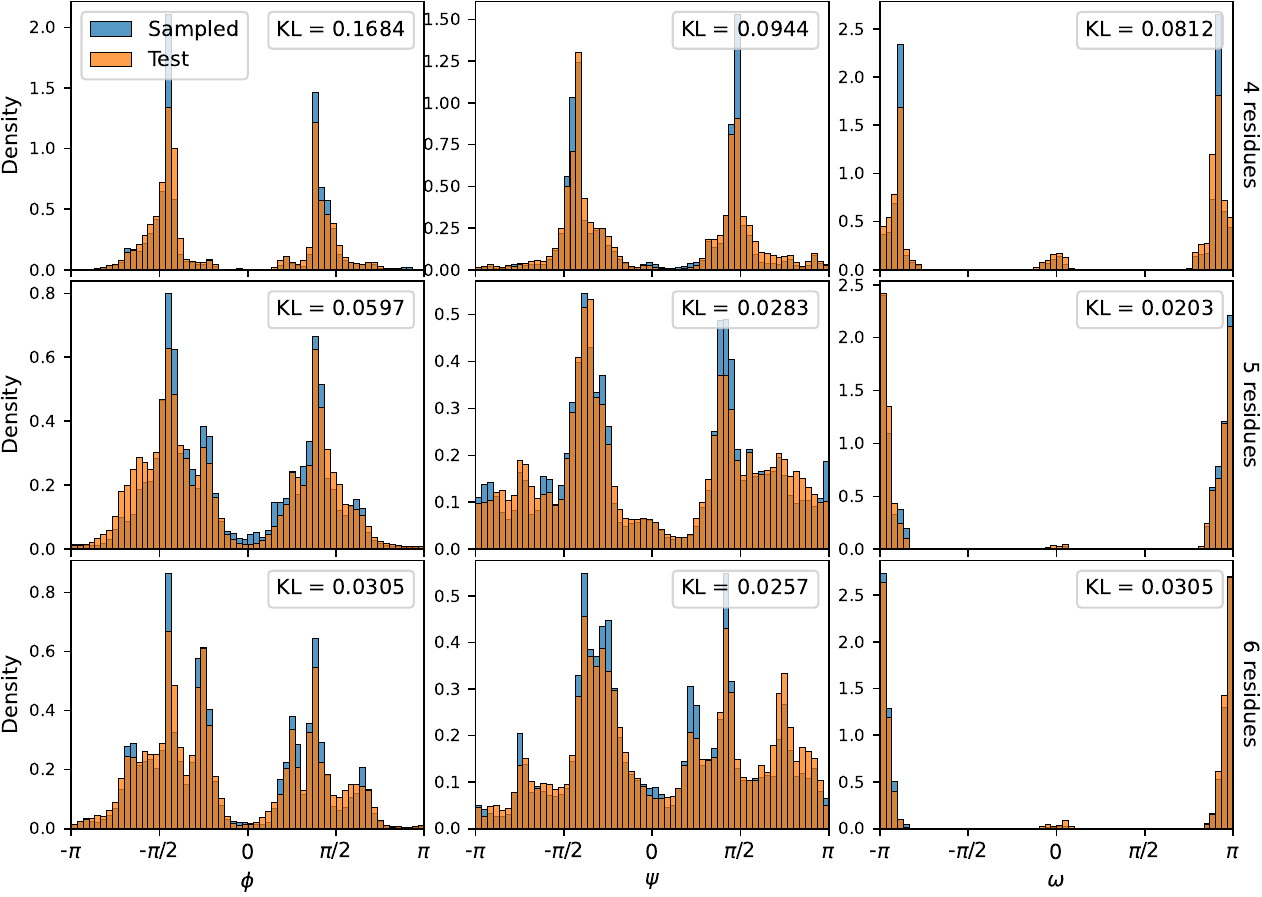}
    \vskip -0.05in
    \caption{Dihedral angle distributions split by number of residues for the backbone-only unconditional model.}
    \label{fig:dihedrals_residues}
\end{figure}
\begin{figure}[H]
    \centering
    \includegraphics[width=0.6\linewidth]{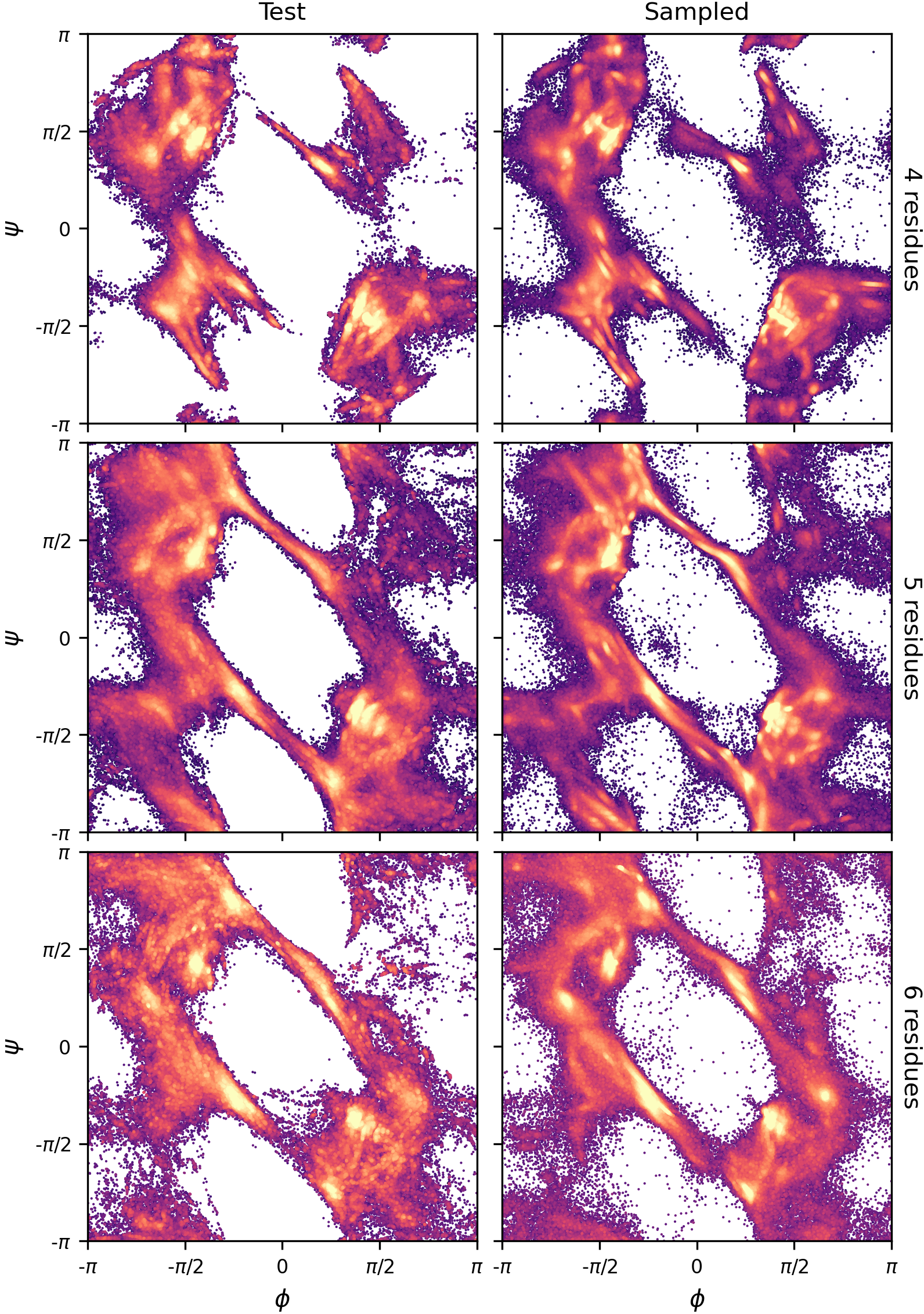}
    \vskip -0.05in
    \caption{Ramachandran distributions split by number of residues for the backbone-only unconditional model.}
    \label{fig:ramas_residues}
\end{figure}

\newpage
\section{Sequence-Conditioned Generation: Backbone-Only Model Results and Plots}
\label{app:add_eval}

We initially investigated sequence-conditioned generation for a backbone-only model to better understand the effect of key hyperparameter choices (e.g. timesteps, number of training set conformers, etc). For completeness, we include the results of these preliminary studies below. 

\begin{table}[H]
    \centering
    \caption{Performance metrics for sequence-conditioned generation of macrocycles evaluated on ring atoms. Coverage is evaluated at a threshold of \SI{0.1}{\angstrom} for rRMSD and 0.05 for rTFD. $k$ is the maximum number of lowest-energy conformers used per molecule in the \emph{training} data. \emph{All test data} conformers are used for evaluation. ``opt'' refers to the use of Equation~\eqref{eq:opt} to reconstruct Cartesian coordinates.}
    \label{tab:rmsd_and_tfd}
    \setlength\tabcolsep{0pt}
    \begin{tabular*}{\linewidth}{@{\extracolsep{\fill}} lccccccccc}
    \toprule
    & & \multicolumn{4}{c}{ \textbf{rRMSD -- Recall} } & \multicolumn{4}{c}{\textbf{rRMSD -- Precision}} \\
    
    \vspace{0.1cm}
    & & \multicolumn{2}{c}{COV (\%) $\uparrow$} & \multicolumn{2}{c}{MAT (\si{\angstrom}) $\downarrow$} \ 
    & \multicolumn{2}{c}{COV (\%) $\uparrow$} & \multicolumn{2}{c}{MAT (\si{\angstrom}) $\downarrow$}\\

    Method & $k$ & Mean & Med. & Mean & Med. & Mean & Med. & Mean & Med. \\
    
    \midrule
    RDKit \cite{Wang2020-ai} & -- & 35.8 & 8.9 & 0.187 & 0.160 & 5.6 & 0.9 & 0.540 & 0.504 \\
    OMEGA \cite{openeyeomega-42} & -- & 32.3 & 7.1 & 0.186 & 0.163 & 3.7 & 1.3 & 0.557 & 0.525 \\
    GeoDiff \cite{xu2022geodiff} & 30 & 50.8 & 54.2 & 0.151 & 0.120 & 6.4 & 3.0 & 0.592 & 0.559 \\
    \textbf{\shortmethodname{}} & 30 & 77.0 & 84.5 & 0.091 & 0.072 & \textbf{61.3} & \textbf{69.1} & \textbf{0.185} & \textbf{0.120} \\
    \textbf{\shortmethodname{}} (opt) & 1 & 63.8 & 66.9 & 0.139 & 0.112 & 58.1 & 65.1 & 0.430 & 0.327 \\
    \textbf{\shortmethodname{}} (opt) & 30 & 79.7 & 86.3 & 0.084 & 0.065 & 56.4 & 62.7 & 0.441 & 0.356 \\
    \textbf{\shortmethodname{}} (opt) & 100 & \textbf{85.6} & \textbf{92.2} & \textbf{0.065} & \textbf{0.049} & 56.9 & 62.4 & 0.454 & 0.385 \\
    \midrule
    & & \multicolumn{4}{c}{ \textbf{rTFD -- Recall} } & \multicolumn{4}{c}{\textbf{rTFD -- Precision}} \\
    
    \vspace{0.1cm}
    & & \multicolumn{2}{c}{COV (\%) $\uparrow$} & \multicolumn{2}{c}{MAT $\downarrow$} \ 
    & \multicolumn{2}{c}{COV (\%) $\uparrow$} & \multicolumn{2}{c}{MAT $\downarrow$}\\

    Method & $k$ & Mean & Med. & Mean & Med. & Mean & Med. & Mean & Med. \\
    \midrule
    RDKit \cite{Wang2020-ai} & -- & 52.9 & 55.3 & 0.059 & 0.051 & 9.4 & 4.4 & 0.215 & 0.206 \\
    OMEGA \cite{openeyeomega-42} & -- & 49.7 & 47.6 & 0.061 & 0.055 & 6.6 & 4.2 & 0.225 & 0.219 \\
    GeoDiff \cite{xu2022geodiff} & 30 & 68.1 & 83.0 & 0.048 & 0.037 & 9.1 & 6.1 & 0.248 & 0.241 \\
    \textbf{\shortmethodname{}} & 30 & \textbf{90.1} & \textbf{95.0} & \textbf{0.024} & \textbf{0.019} & \textbf{74.7} & \textbf{86.2} & \textbf{0.059} & \textbf{0.033} \\
    \textbf{\shortmethodname{}} (opt) & 30 & 89.2 & 94.3 & \textbf{0.024} & \textbf{0.019} & 61.8 & 68.9 & 0.068 & 0.044 \\
    \bottomrule
    \end{tabular*}%
\end{table}%
\begin{table}[H]
    \centering
    \caption{Evaluating RINGER (opt) trained and sampled with different numbers of timesteps.}
    \label{tab:timesteps}
    \begin{tabular}{lcccccccc}
    \toprule
    \multicolumn{0}{c}{  } & \multicolumn{4}{c}{ \textbf{rRMSD -- Recall} } & \multicolumn{4}{c}{\textbf{rRMSD -- Precision}} \\
    
    \vspace{0.1cm}
    & \multicolumn{2}{c}{COV (\%) $\uparrow$} & \multicolumn{2}{c}{MAT (\si{\angstrom}) $\downarrow$} \ 
    & \multicolumn{2}{c}{COV (\%) $\uparrow$} & \multicolumn{2}{c}{MAT (\si{\angstrom}) $\downarrow$}\\

    Timesteps & Mean & Med. & Mean & Med. & Mean & Med. & Mean & Med. \\
    
    \midrule
    20 & 79.7 & 86.3 & 0.084 & 0.065 & 56.4 & 62.7 & 0.441 & 0.356 \\
    50 & 80.8 & 88.0 & 0.082 & 0.061 & \textbf{60.5} & \textbf{68.9} & \textbf{0.431} & \textbf{0.335} \\
    100 & \textbf{81.5} & \textbf{88.9} & \textbf{0.080} & \textbf{0.060} & 58.0 & 65.2 & 0.443 & 0.365 \\
    \bottomrule
    \end{tabular}%
\end{table}%

Table~\ref{tab:dihedrals} shows that bond angles are required in addition to dihedral angles in order for the model to perform well. To reconstruct Cartesian geometries using the dihedral-only model, we modified Equation~\eqref{eq:opt} to include inequality constraints for the bond angles where the upper and lower limit are determined by the standard deviations of bond angles from the training data.

\begin{table}[H]
    \centering
    \caption{Evaluating RINGER (opt) trained only with dihedral angles.}
    \label{tab:dihedrals}
    \begin{tabular}{lcccccccc}
    \toprule
    \multicolumn{0}{c}{  } & \multicolumn{4}{c}{ \textbf{rRMSD -- Recall} } & \multicolumn{4}{c}{\textbf{rRMSD -- Precision}} \\
    
    \vspace{0.1cm}
    & \multicolumn{2}{c}{COV (\%) $\uparrow$} & \multicolumn{2}{c}{MAT (\si{\angstrom}) $\downarrow$} \ 
    & \multicolumn{2}{c}{COV (\%) $\uparrow$} & \multicolumn{2}{c}{MAT (\si{\angstrom}) $\downarrow$}\\

     & Mean & Med. & Mean & Med. & Mean & Med. & Mean & Med. \\
    
    \midrule
    $\mathbf{x}_i=[\theta_i,\tau_i]$ & \textbf{79.7} & \textbf{86.3} & \textbf{0.084} & \textbf{0.065} & \textbf{56.4} & \textbf{62.7} & \textbf{0.441} & \textbf{0.356} \\
    $\mathbf{x}_i=[\tau_i]$ & 66.8 & 73.5 & 0.130 & 0.101 & 41.1 & 37.6 & 0.469 & 0.417 \\
    \midrule
    \multicolumn{0}{c}{  } & \multicolumn{4}{c}{ \textbf{rTFD -- Recall} } & \multicolumn{4}{c}{\textbf{rTFD -- Precision}} \\
    
    \vspace{0.1cm}
    & \multicolumn{2}{c}{COV (\%) $\uparrow$} & \multicolumn{2}{c}{MAT $\downarrow$} \ 
    & \multicolumn{2}{c}{COV (\%) $\uparrow$} & \multicolumn{2}{c}{MAT $\downarrow$}\\

     & Mean & Med. & Mean & Med. & Mean & Med. & Mean & Med. \\
    
    \midrule
    $\mathbf{x}_i=[\theta_i,\tau_i]$ & \textbf{89.2} & \textbf{94.3} & \textbf{0.024} & \textbf{0.019} & \textbf{61.8} & \textbf{68.9} & \textbf{0.068} & \textbf{0.044} \\
    $\mathbf{x}_i=[\tau_i]$ & 83.2 & 91.0 & 0.035 & 0.024 & 49.2 & 49.3 & 0.144 & 0.114 \\
    \bottomrule
    \end{tabular}%
\end{table}%

\newpage
\section{Sequence-Conditioned Generation: Additional Results and Plots}
\label{app:cond}

\subsection{Distributions}
\label{app:dists}

\begin{figure}[H]
    \centering
    \begin{minipage}{0.71\linewidth}
        \includegraphics[width=\linewidth]{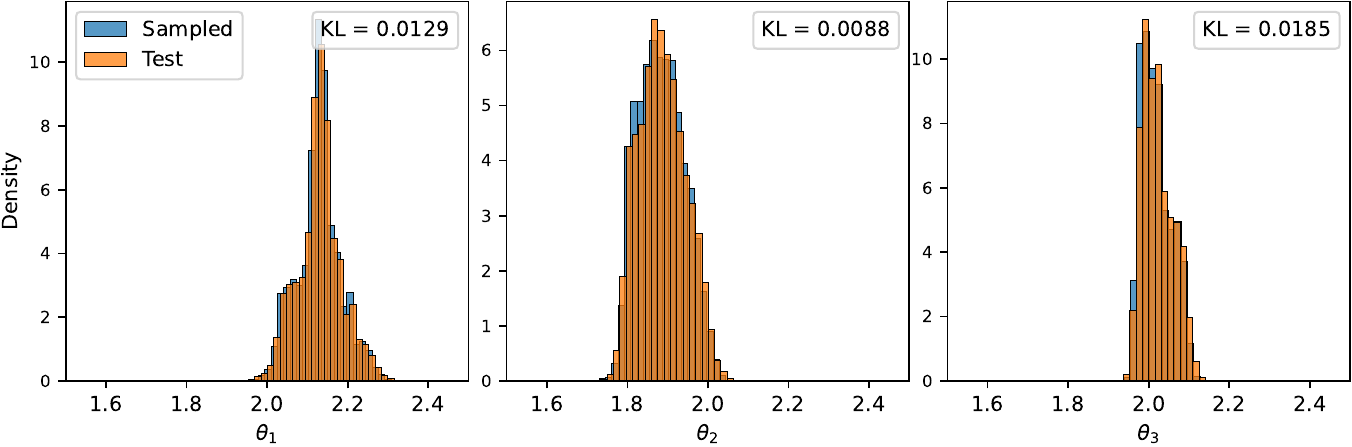}
        \includegraphics[width=\linewidth]{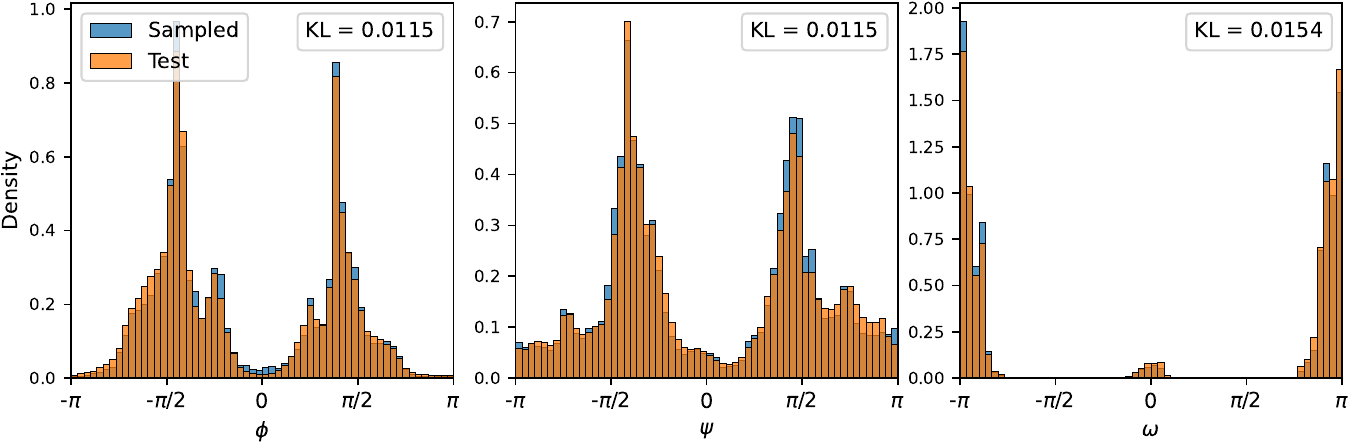}
    \end{minipage}\hfill%
    \begin{minipage}{0.275\linewidth}
        \includegraphics[width=\linewidth]{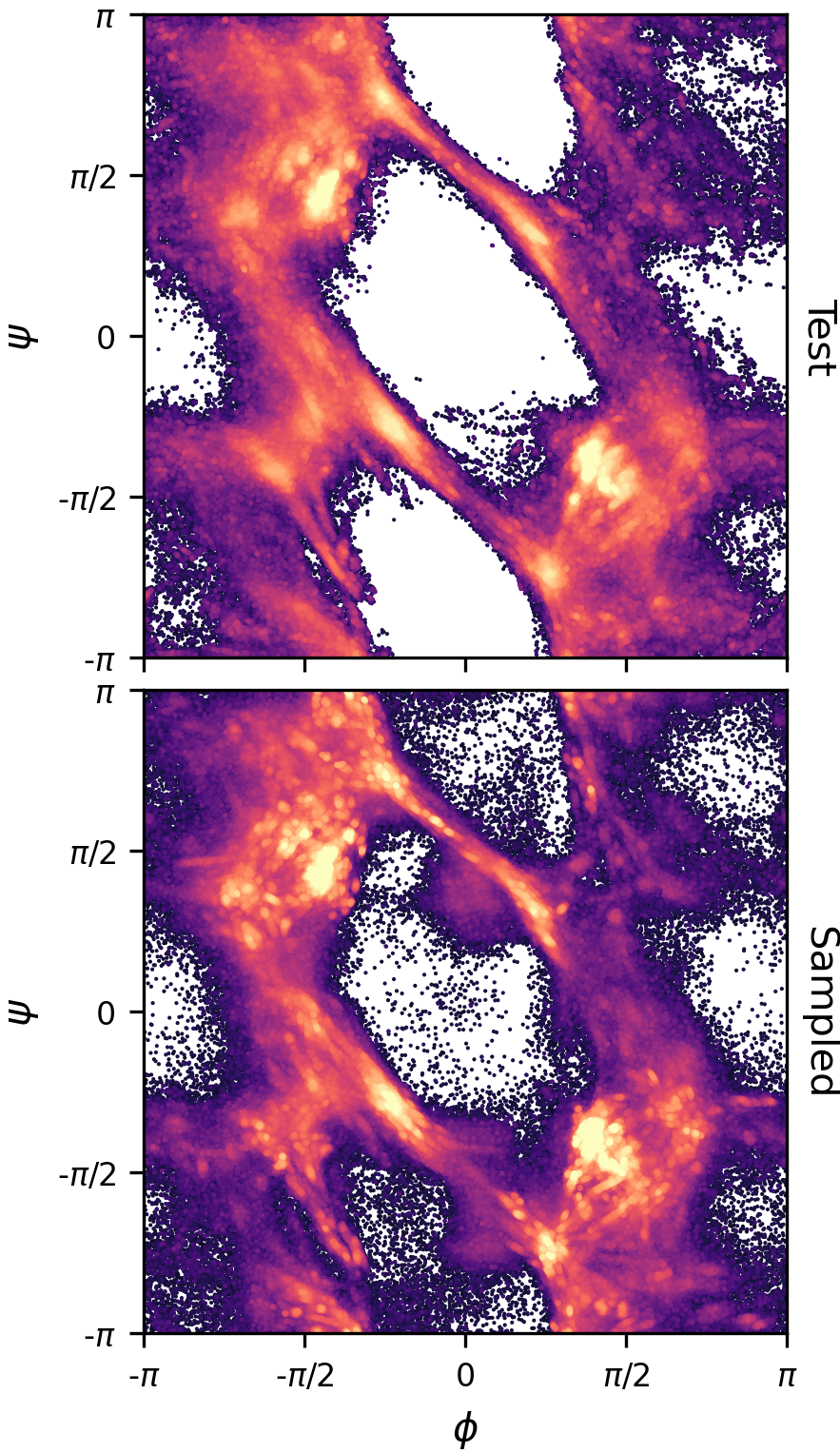}
    \end{minipage}
    \includegraphics[width=\linewidth]{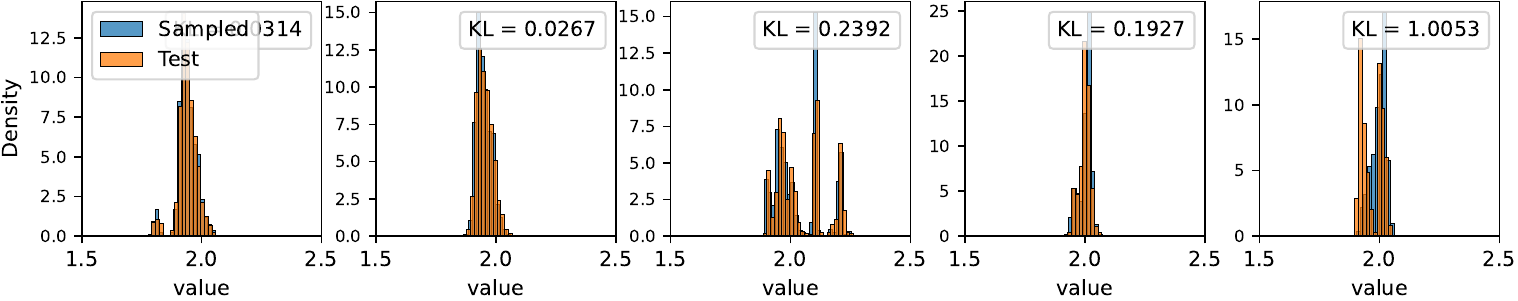}
    \includegraphics[width=\linewidth]{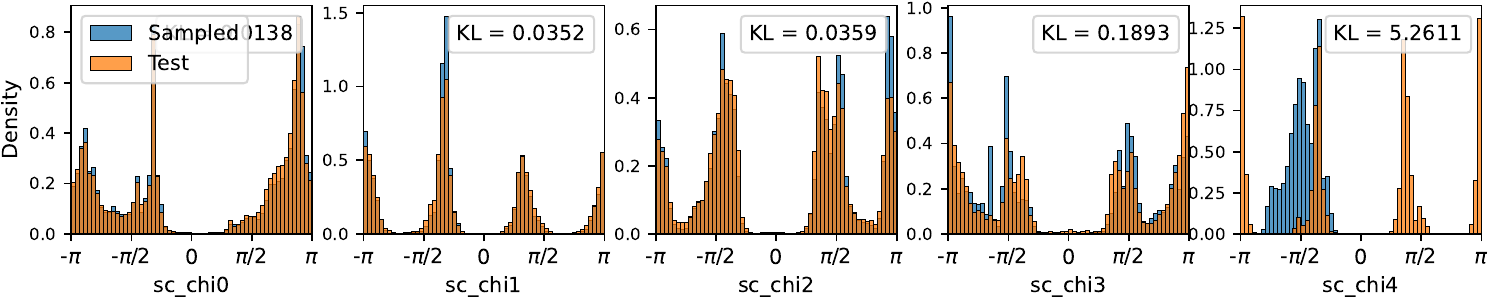}
    \vskip -0.05in
    \caption{Comparison of the bond angle and dihedral distributions from the held-out test set and in the \emph{conditionally} generated samples (prior to reconstruction to Cartesian coordinates). The top two rows show the distributions of internal coordinates in the ring and the bottom two rows show the side-chain distributions.}
    \label{fig:dists_cond}
\end{figure}

\begin{figure}[H]
    \centering
    \includegraphics[width=0.8\linewidth]{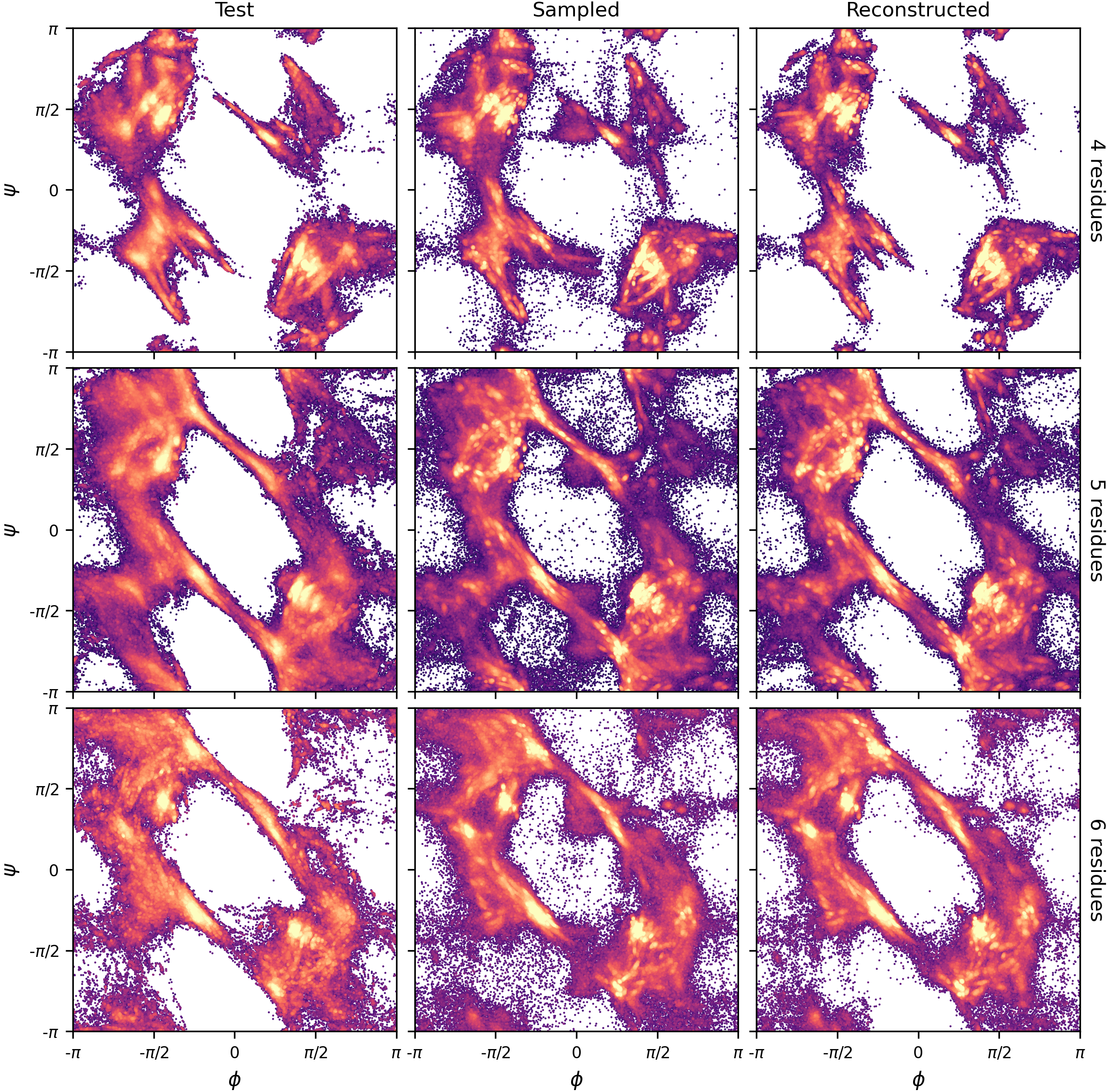}
    \vskip -0.05in
    \caption{Ramachandran distributions for \emph{conditionally} generated samples split by number of residues. The ``Reconstructed'' column shows the distributions after converting to Cartesian coordinates using the SLSQP optimization in Equation~\eqref{eq:opt} followed by rejection of inadequate samples.}
    \label{fig:ramas_cond}
\end{figure}

Figure~\ref{fig:ramas_cond} shows the Ramachandran plots split by number of residues for the conditional model and illustrates the effect of Equation~\eqref{eq:opt} to reconstruct realizable Cartesian geometries from the set of redundant internal coordinates predicted by the model. Notably, while the reconstructed geometries still reproduce the joint distribution over dihedral angles well, several artifacts are introduced as a result of the optimization, which motivates the further development of generative methods that directly incorporate the cyclic constraints into the diffusion process itself.

\subsection{Coverage}
\label{app:coverage}

\begin{figure}[H]
    \centering
    \includegraphics[width=\linewidth]{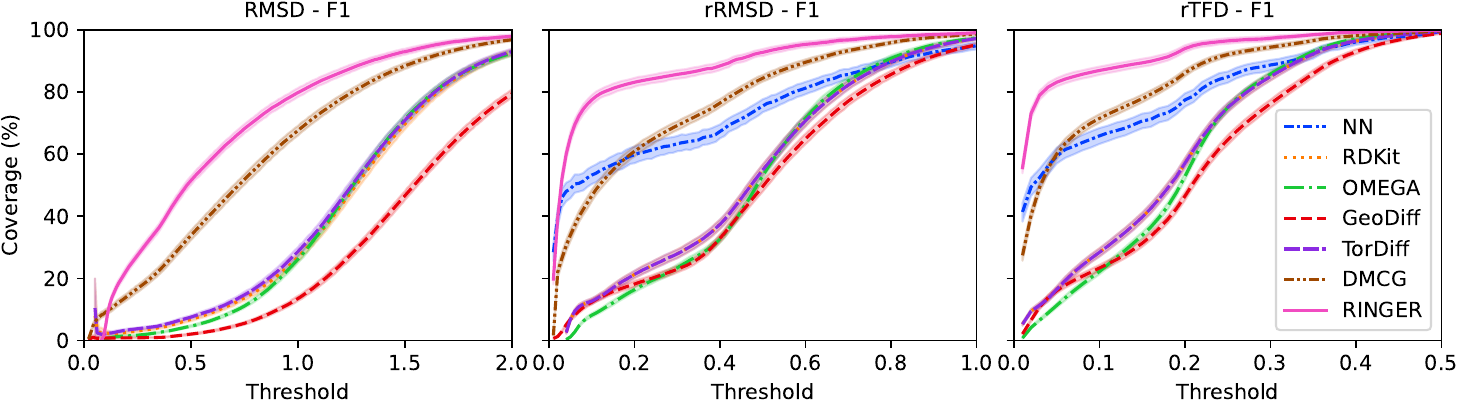}
    \vskip -0.05in
    \caption{Comparison of mean coverage when varying the threshold. Translucent error bands correspond to 95\% confidence intervals.}
    \label{fig:cov}
\end{figure}

\subsection{Tables of Performance Metrics}

\begin{table}[H]
    \centering
    \caption{Performance metrics for sequence-conditioned generation of macrocycles. Coverage is evaluated at a threshold of \SI{0.75}{\angstrom} for all-atom RMSD, \SI{0.1}{\angstrom} for ring-only RMSD (rRMSD), and 0.05 for ring-only TFD (rTFD). \emph{All test data} conformers are used for evaluation.}
    \label{tab:metrics_all_atom}
    \setlength\tabcolsep{0pt}
    \begin{tabular*}{\linewidth}{@{\extracolsep{\fill}} lcccccccccccc}
    \toprule
    & \multicolumn{4}{c}{ \textbf{RMSD -- Recall} } & \multicolumn{4}{c}{\textbf{RMSD -- Precision}} & \multicolumn{4}{c}{ \textbf{RMSD -- F1} } \\
    
    & \multicolumn{2}{c}{COV (\%) $\uparrow$} & \multicolumn{2}{c}{MAT (\si{\angstrom}) $\downarrow$} & \multicolumn{2}{c}{COV (\%) $\uparrow$} & \multicolumn{2}{c}{MAT (\si{\angstrom}) $\downarrow$} & \multicolumn{2}{c}{COV (\%) $\uparrow$} & \multicolumn{2}{c}{MAT (\si{\angstrom}) $\downarrow$} \\

    Method & Mean & Med. & Mean & Med. & Mean & Med. & Mean & Med. & Mean & Med. & Mean & Med. \\
    
    \midrule

RDKit \cite{Wang2020-ai} & 41.1 & 33.0 & 0.853 & 0.830 & 7.5 & 3.7 & 1.419 & 1.357 & 12.7 & 6.7 & 1.065 & 1.030 \\
OMEGA \cite{openeyeomega-42} & 36.2 & 31.3 & 0.900 & 0.852 & 6.4 & 4.9 & 1.433 & 1.360 & 10.9 & 8.5 & 1.106 & 1.047 \\
GeoDiff \cite{xu2022geodiff} & 25.6 & 14.3 & 0.989 & 0.949 & 2.7 & 1.1 & 1.753 & 1.701 & 4.9 & 2.0 & 1.265 & 1.218 \\
TorDiff \cite{Jing2022-torsionaldiffusion} & 44.4 & 38.5 & 0.820 & 0.797 & 8.3 & 4.5 & 1.399 & 1.338 & 14.0 & 8.1 & 1.034 & 0.999 \\
DMCG \cite{zhu2022dmcg} & \textbf{78.3} & \textbf{88.0} & \textbf{0.534} & \textbf{0.496} & 41.9 & 40.6 & 0.942 & 0.889 & 54.6 & 55.6 & 0.682 & 0.636 \\
\textbf{\shortmethodname{}} & 62.3 & 64.6 & 0.691 & 0.636 & \textbf{79.6} & \textbf{95.2} & \textbf{0.468} & \textbf{0.319} & \textbf{69.9} & \textbf{76.9} & \textbf{0.558} & \textbf{0.425} \\
    
    \midrule
    
    & \multicolumn{4}{c}{ \textbf{rRMSD -- Recall} } & \multicolumn{4}{c}{\textbf{rRMSD -- Precision}} & \multicolumn{4}{c}{ \textbf{rRMSD -- F1} } \\
    & \multicolumn{2}{c}{COV (\%) $\uparrow$} & \multicolumn{2}{c}{MAT (\si{\angstrom}) $\downarrow$} & \multicolumn{2}{c}{COV (\%) $\uparrow$} & \multicolumn{2}{c}{MAT (\si{\angstrom}) $\downarrow$} & \multicolumn{2}{c}{COV (\%) $\uparrow$} & \multicolumn{2}{c}{MAT (\si{\angstrom}) $\downarrow$} \\
    
    \midrule

1-NN & 43.7 & 35.5 & 0.301 & 0.182 & 40.3 & 20.6 & 0.331 & 0.244 & 41.9 & 26.1 & 0.315 & 0.208 \\
RDKit \cite{Wang2020-ai} & 35.8 & 8.9 & 0.187 & 0.160 & 5.6 & 0.9 & 0.540 & 0.504 & 9.7 & 1.6 & 0.277 & 0.243 \\
OMEGA \cite{openeyeomega-42} & 32.2 & 7.1 & 0.186 & 0.163 & 3.7 & 1.3 & 0.557 & 0.525 & 6.6 & 2.2 & 0.279 & 0.249 \\
GeoDiff \cite{xu2022geodiff} & 50.8 & 54.2 & 0.151 & 0.120 & 6.4 & 3.0 & 0.592 & 0.559 & 11.4 & 5.7 & 0.240 & 0.198 \\
TorDiff \cite{Jing2022-torsionaldiffusion} & 35.8 & 8.9 & 0.187 & 0.160 & 5.6 & 0.9 & 0.540 & 0.504 & 9.7 & 1.6 & 0.277 & 0.243 \\
DMCG \cite{zhu2022dmcg} & \textbf{77.6} & \textbf{89.1} & \textbf{0.076} & \textbf{0.061} & 36.7 & 32.7 & 0.301 & 0.260 & 49.8 & 47.8 & 0.121 & 0.099 \\
\textbf{\shortmethodname{}} & 76.3 & 82.6 & 0.110 & 0.077 & \textbf{81.6} & \textbf{96.6} & \textbf{0.108} & \textbf{0.037} & \textbf{78.9} & \textbf{89.1} & \textbf{0.109} & \textbf{0.050} \\
    
    \midrule
    
    & \multicolumn{4}{c}{ \textbf{rTFD -- Recall} } & \multicolumn{4}{c}{\textbf{rTFD -- Precision}} & \multicolumn{4}{c}{ \textbf{rTFD -- F1} } \\
    & \multicolumn{2}{c}{COV (\%) $\uparrow$} & \multicolumn{2}{c}{MAT $\downarrow$} & \multicolumn{2}{c}{COV (\%) $\uparrow$} & \multicolumn{2}{c}{MAT $\downarrow$} & \multicolumn{2}{c}{COV (\%) $\uparrow$} & \multicolumn{2}{c}{MAT $\downarrow$} \\
    
    \midrule
    
1-NN & 53.1 & 67.7 & 0.111 & 0.054 & 48.6 & 51.1 & 0.122 & 0.078 & 50.7 & 58.2 & 0.116 & 0.064 \\
RDKit \cite{Wang2020-ai} & 52.9 & 55.3 & 0.059 & 0.051 & 9.4 & 4.4 & 0.215 & 0.206 & 15.9 & 8.2 & 0.093 & 0.082 \\
OMEGA \cite{openeyeomega-42} & 49.7 & 47.6 & 0.061 & 0.055 & 6.6 & 4.2 & 0.225 & 0.219 & 11.7 & 7.7 & 0.095 & 0.088 \\
GeoDiff \cite{xu2022geodiff} & 68.1 & 83.0 & 0.048 & 0.037 & 9.1 & 6.1 & 0.248 & 0.241 & 16.0 & 11.4 & 0.080 & 0.064 \\
TorDiff \cite{Jing2022-torsionaldiffusion} & 52.9 & 55.3 & 0.059 & 0.051 & 9.4 & 4.4 & 0.215 & 0.206 & 15.9 & 8.2 & 0.093 & 0.082 \\
DMCG \cite{zhu2022dmcg} & \textbf{93.0} & \textbf{98.2} & \textbf{0.021} & \textbf{0.017} & 48.5 & 50.6 & 0.110 & 0.097 & 63.8 & 66.8 & \textbf{0.036} & 0.029 \\
\textbf{\shortmethodname{}} & 83.8 & 90.2 & 0.035 & 0.024 & \textbf{85.3} & \textbf{98.4} & \textbf{0.038} & \textbf{0.011} & \textbf{84.5} & \textbf{94.1} & 0.037 & \textbf{0.015} \\
    
    \bottomrule
    \end{tabular*}%
\end{table}%

\subsection{Cyclic Positional Encoding Ablation Study}
\label{app:cyc_ablation}

To assess the impact of the cyclic relative positional encoding in Equations~\eqref{eq:cyclic1} and \eqref{eq:cyclic2}, we trained two models on 10\% of the training data for 100 epochs: one with a standard relative positional encoding and one with our cyclic relative positional encoding. The results in Tables~\ref{tab:cyc_ablation} and \ref{tab:cyc_ablation_residues} illustrate how our newly designed encoding improves performance, especially for larger macrocycles.

\begin{table}[H]
    \centering
    \caption{Ablation study comparing \shortmethodname{} performance with and without cyclic relative positional encoding defined in \eqref{eq:cyclic1} and \eqref{eq:cyclic2}. Models trained on 10\% of data for 100 epochs.}
    \label{tab:cyc_ablation}
    \setlength\tabcolsep{0pt}
    \begin{tabular*}{\linewidth}{@{\extracolsep{\fill}} lcccccccc}
    \toprule
    \multicolumn{0}{c}{  } & \multicolumn{4}{c}{ \textbf{RMSD -- Recall} } & \multicolumn{4}{c}{\textbf{RMSD -- Precision}} \\
    & \multicolumn{2}{c}{COV (\%) $\uparrow$} & \multicolumn{2}{c}{MAT (\si{\angstrom}) $\downarrow$} \ 
    & \multicolumn{2}{c}{COV (\%) $\uparrow$} & \multicolumn{2}{c}{MAT (\si{\angstrom}) $\downarrow$}\\
     & Mean & Med. & Mean & Med. & Mean & Med. & Mean & Med. \\
    
    \midrule

    Standard encoding & 33.9 & 26.2 & 0.913 & 0.872 & 10.1 & 3.8 & 1.610 & 1.560 \\
    Cyclic encoding & \textbf{40.4} & \textbf{38.6} & \textbf{0.861} & \textbf{0.819} & \textbf{14.2} & \textbf{7.4} & \textbf{1.538} & \textbf{1.491} \\
    \midrule
    \multicolumn{0}{c}{  } & \multicolumn{4}{c}{ \textbf{rRMSD – Recall} } & \multicolumn{4}{c}{\textbf{rRMSD – Precision}} \\
    & \multicolumn{2}{c}{COV (\%) $\uparrow$} & \multicolumn{2}{c}{MAT (\si{\angstrom}) $\downarrow$} \ 
    & \multicolumn{2}{c}{COV (\%) $\uparrow$} & \multicolumn{2}{c}{MAT (\si{\angstrom}) $\downarrow$}\\
    \midrule
    Standard encoding & 26.2 & 1.387 & 0.213 & 0.191 & 5.8 & 0.1 & 0.721 & 0.701 \\ 
    Cyclic encoding & \textbf{32.3} & \textbf{13.7} & \textbf{0.189} & \textbf{0.166} & \textbf{9.6} & \textbf{0.8} & \textbf{0.678} & \textbf{0.658} \\ 
    \midrule       
    \multicolumn{0}{c}{  } & \multicolumn{4}{c}{ \textbf{rTFD – Recall} } & \multicolumn{4}{c}{\textbf{rTFD – Precision}} \\
    & \multicolumn{2}{c}{COV (\%) $\uparrow$} & \multicolumn{2}{c}{MAT $\downarrow$} \ 
    & \multicolumn{2}{c}{COV (\%) $\uparrow$} & \multicolumn{2}{c}{MAT $\downarrow$}\\
    \midrule
    Standard encoding & 47.2 & 50.0 & 0.064 & 0.055 & 11.0 & 5.1 & 0.168 & 0.149 \\
    Cyclic encoding & \textbf{55.6} & \textbf{62.9} & \textbf{0.056} & \textbf{0.048} & \textbf{16.7} & \textbf{10.9} & \textbf{0.154} & \textbf{0.134} \\
    
    \bottomrule
    \end{tabular*}%
\end{table}%

\begin{table}[H]
    \centering
    \caption{Ablation study comparing \shortmethodname{} rTFD performance with and without cyclic relative positional encoding across different macrocycle sizes. Models trained on 10\% of data for 100 epochs.}
    \label{tab:cyc_ablation_residues}
    \setlength\tabcolsep{0pt}
    \begin{tabular*}{\linewidth}{@{\extracolsep{\fill}} lccccccccc}
    \toprule
    & & \multicolumn{4}{c}{ \textbf{rTFD -- Recall} } & \multicolumn{4}{c}{\textbf{rTFD -- Precision}} \\
    & & \multicolumn{2}{c}{COV (\%) $\uparrow$} & \multicolumn{2}{c}{MAT $\downarrow$}  & \multicolumn{2}{c}{COV (\%) $\uparrow$} & \multicolumn{2}{c}{MAT $\downarrow$} \\
    & \#residues & Mean & Med. & Mean & Med. & Mean & Med. & Mean & Med. \\

    \midrule

    Standard encoding & 4 & 62.2 & 76.5 & 0.056 & 0.038 & 18.0 & 20.1 & 0.175 & 0.143 \\
    Cyclic encoding & 4 & \textbf{69.1} & \textbf{84.0} & \textbf{0.047} & \textbf{0.032} & \textbf{25.3} & \textbf{29.6} & \textbf{0.160} & \textbf{0.128} \\
    \midrule
    Standard encoding & 5 & 37.1 & 35.6 & 0.066 & 0.062 & 4.1 & 3.0 & 0.151 & 0.141 \\
    Cyclic encoding & 5 & \textbf{47.0} & \textbf{51.4} & \textbf{0.061} & \textbf{0.055} & \textbf{9.0} & \textbf{7.5} & \textbf{0.137} & \textbf{0.127} \\
    \midrule       
    Standard encoding & 6 & 9.8 & 1.2 & 0.090 & 0.080 & 0.3 & 0.1 & 0.186 & 0.178 \\
    Cyclic encoding & 6 & \textbf{20.4} & \textbf{5.1} & \textbf{0.079} & \textbf{0.071} & \textbf{0.8} & \textbf{0.4} & \textbf{0.177} & \textbf{0.169} \\
    
    \bottomrule
    \end{tabular*}%
\end{table}%

\section{Post Hoc Optimization with GFN2-xTB}
\label{app:xtb_opt}

To further evaluate the sampling performance as well as macrocycle conformer quality we performed post hoc optimization with GFN2-xTB. This provides a level comparison between sampling methods. As shown in Table~\ref{tab:metrics_xtb} below, \shortmethodname{} maintains excellent performance across recall, precision, and F1 metrics across both all-atom and backbone-only evaluations.  

\begin{table}[H]
    \centering
    \caption{Performance metrics for samples \emph{with post hoc optimization using GFN2-xTB}. Coverage is evaluated at a threshold of \SI{0.75}{\angstrom} for all-atom RMSD, \SI{0.1}{\angstrom} for ring-only RMSD (rRMSD), and 0.05 for ring-only TFD (rTFD). \emph{All test data} conformers are used for evaluation.}
    \label{tab:metrics_xtb}
    \setlength\tabcolsep{0pt}
    \begin{tabular*}{\linewidth}{@{\extracolsep{\fill}} lcccccccccccc}
    \toprule
    & \multicolumn{4}{c}{ \textbf{RMSD -- Recall} } & \multicolumn{4}{c}{\textbf{RMSD -- Precision}} & \multicolumn{4}{c}{ \textbf{RMSD -- F1} } \\
    
    & \multicolumn{2}{c}{COV (\%) $\uparrow$} & \multicolumn{2}{c}{MAT (\si{\angstrom}) $\downarrow$} & \multicolumn{2}{c}{COV (\%) $\uparrow$} & \multicolumn{2}{c}{MAT (\si{\angstrom}) $\downarrow$} & \multicolumn{2}{c}{COV (\%) $\uparrow$} & \multicolumn{2}{c}{MAT (\si{\angstrom}) $\downarrow$} \\

    Method & Mean & Med. & Mean & Med. & Mean & Med. & Mean & Med. & Mean & Med. & Mean & Med. \\
    
    \midrule

RDKit \cite{Wang2020-ai} (xTB) & 53.9 & 60.1 & 0.734 & 0.688 & 10.3 & 7.0 & 1.365 & 1.311 & 17.2 & 12.6 & 0.955 & 0.903 \\
OMEGA \cite{openeyeomega-42} (xTB) & 44.9 & 45.0 & 0.818 & 0.769 & 8.7 & 7.1 & 1.392 & 1.317 & 14.5 & 12.2 & 1.031 & 0.971 \\
GeoDiff \cite{xu2022geodiff} (xTB) & 29.9 & 21.4 & 0.938 & 0.896 & 3.1 & 1.5 & 1.702 & 1.652 & 5.7 & 2.9 & 1.209 & 1.162 \\
TorDiff \cite{Jing2022-torsionaldiffusion} (xTB) & 62.5 & 76.9 & 0.641 & 0.570 & 12.5 & 9.7 & 1.327 & 1.267 & 20.9 & 17.2 & 0.865 & 0.786 \\
DMCG \cite{zhu2022dmcg} (xTB) & \textbf{84.9} & \textbf{92.2} & \textbf{0.415} & \textbf{0.371} & 47.4 & 50.0 & 0.856 & 0.788 & 60.8 & 64.8 & 0.559 & 0.505 \\
\textbf{\shortmethodname{}} (xTB) & 63.7 & 66.4 & 0.653 & 0.595 & \textbf{80.6} & \textbf{95.8} & \textbf{0.380} & \textbf{0.216} & \textbf{71.2} & \textbf{78.4} & \textbf{0.481} & \textbf{0.317} \\
    
    \midrule
    
    & \multicolumn{4}{c}{ \textbf{rRMSD -- Recall} } & \multicolumn{4}{c}{\textbf{rRMSD -- Precision}} & \multicolumn{4}{c}{ \textbf{rRMSD -- F1} } \\
    & \multicolumn{2}{c}{COV (\%) $\uparrow$} & \multicolumn{2}{c}{MAT (\si{\angstrom}) $\downarrow$} & \multicolumn{2}{c}{COV (\%) $\uparrow$} & \multicolumn{2}{c}{MAT (\si{\angstrom}) $\downarrow$} & \multicolumn{2}{c}{COV (\%) $\uparrow$} & \multicolumn{2}{c}{MAT (\si{\angstrom}) $\downarrow$} \\
    
    \midrule

1-NN & 43.7 & 35.5 & 0.301 & 0.182 & 40.3 & 20.6 & 0.331 & 0.244 & 41.9 & 26.1 & 0.315 & 0.208 \\
RDKit \cite{Wang2020-ai} (xTB) & 73.0 & 91.2 & 0.098 & 0.057 & 13.4 & 9.8 & 0.508 & 0.466 & 22.7 & 17.6 & 0.164 & 0.102 \\
OMEGA \cite{openeyeomega-42} (xTB) & 68.3 & 82.5 & 0.102 & 0.070 & 10.5 & 9.0 & 0.534 & 0.501 & 18.1 & 16.2 & 0.171 & 0.123 \\
GeoDiff \cite{xu2022geodiff} (xTB) & 64.0 & 75.1 & 0.119 & 0.086 & 8.8 & 5.6 & 0.573 & 0.541 & 15.5 & 10.4 & 0.197 & 0.148 \\
TorDiff \cite{Jing2022-torsionaldiffusion} (xTB) & 81.4 & 95.9 & 0.075 & 0.041 & 16.8 & 14.3 & 0.495 & 0.454 & 27.9 & 25.0 & 0.130 & 0.075 \\
DMCG \cite{zhu2022dmcg} (xTB) & \textbf{93.2} & \textbf{97.6} & \textbf{0.039} & \textbf{0.029} & 53.2 & 57.2 & 0.270 & 0.220 & 67.8 & 72.1 & \textbf{0.068} & 0.051 \\
\textbf{\shortmethodname{}} (xTB) & 78.0 & 84.3 & 0.105 & 0.071 & \textbf{83.3} & \textbf{97.2} & \textbf{0.100} & \textbf{0.027} & \textbf{80.6} & \textbf{90.3} & 0.102 & \textbf{0.039} \\
    
    \midrule
    
    & \multicolumn{4}{c}{ \textbf{rTFD -- Recall} } & \multicolumn{4}{c}{\textbf{rTFD -- Precision}} & \multicolumn{4}{c}{ \textbf{rTFD -- F1} } \\
    & \multicolumn{2}{c}{COV (\%) $\uparrow$} & \multicolumn{2}{c}{MAT $\downarrow$} & \multicolumn{2}{c}{COV (\%) $\uparrow$} & \multicolumn{2}{c}{MAT $\downarrow$} & \multicolumn{2}{c}{COV (\%) $\uparrow$} & \multicolumn{2}{c}{MAT $\downarrow$} \\
    
    \midrule
    
1-NN & 53.1 & 67.7 & 0.111 & 0.054 & 48.6 & 51.1 & 0.122 & 0.078 & 50.7 & 58.2 & 0.116 & 0.064 \\
RDKit \cite{Wang2020-ai} (xTB) & 85.6 & 98.3 & 0.029 & 0.017 & 17.3 & 14.5 & 0.200 & 0.189 & 28.8 & 25.3 & 0.051 & 0.031 \\
OMEGA \cite{openeyeomega-42} (xTB) & 83.8 & 95.8 & 0.029 & 0.021 & 13.4 & 11.7 & 0.212 & 0.204 & 23.1 & 20.8 & 0.051 & 0.038 \\
GeoDiff \cite{xu2022geodiff} (xTB) & 79.8 & 94.0 & 0.037 & 0.026 & 11.7 & 9.3 & 0.242 & 0.233 & 20.4 & 17.0 & 0.065 & 0.047 \\
TorDiff \cite{Jing2022-torsionaldiffusion} (xTB) & 89.8 & 99.4 & 0.023 & 0.012 & 20.0 & 17.8 & 0.195 & 0.184 & 32.8 & 30.2 & 0.041 & 0.023 \\
DMCG \cite{zhu2022dmcg} (xTB) & \textbf{96.9} & \textbf{99.6} & \textbf{0.012} & \textbf{0.008} & 57.5 & 62.1 & 0.099 & 0.085 & 72.2 & 76.5 & \textbf{0.021} & 0.015 \\
\textbf{\shortmethodname{}} (xTB) & 84.1 & 90.4 & 0.034 & 0.022 & \textbf{85.5} & \textbf{98.2} & \textbf{0.036} & \textbf{0.008} & \textbf{84.8} & \textbf{94.1} & 0.035 & \textbf{0.012} \\
    
    \bottomrule
    \end{tabular*}%
\end{table}%

\begin{figure}[H]
    \centering
    \includegraphics[width=0.8\linewidth]{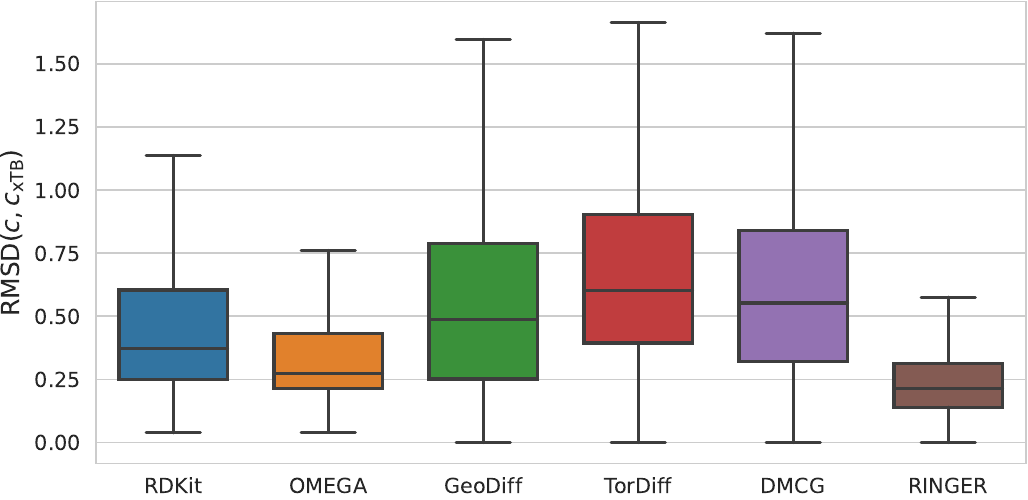}
    \vskip -0.05in
    \caption{RMSD of generated conformers before and after xTB optimization. On average, \shortmethodname{}-generated samples require less structural modification to reach xTB local minima. Outliers are not shown for clarity.}
    \label{fig:rmsd_xtb_changes}
\end{figure}

\end{document}